\documentclass[11pt,a4paper]{article}
 
\usepackage{amsmath,amsthm,amssymb}

\pdfoutput=1

\usepackage{graphics,graphicx}
\usepackage{epsfig}
\usepackage{multicol}
\usepackage{color}
\makeatletter
\@addtoreset{equation}{section}
\makeatother

\usepackage{braket}
\usepackage{bm}

\begin{document} 
\sloppy

\begin{center}
\LARGE{{\bf A New Approach to Generalised Uncertainty Relations}}
\end{center}

\begin{center}
\large{Matthew J. Lake} ${}^{a,b,c,d}$\footnote{matthewjlake@narit.or.th}
\end{center}
\begin{center}
\emph{ ${}^a$ National Astronomical Research Institute of Thailand, \\ 260 Moo 4, T. Donkaew, A. Maerim, Chiang Mai 50180, Thailand \\}
\emph{ ${}^b$ Department of Physics, Faculty of Science, Chiang Mai University, \\ 239 Huaykaew Road, T. Suthep, A. Muang, Chiang Mai 50200, Thailand \\}
\emph{ ${}^c$ School of Physics, Sun Yat-Sen University, \\ Guangzhou 510275, People's Republic of China, \\}
\emph{ ${}^d$ Department of Physics, Babe\c s-Bolyai University, \\ Mihail Kog\u alniceanu Street 1, 400084 Cluj-Napoca, Romania \\} 
\vspace{0.1cm}
\end{center}

\begin{abstract}
We outline a new model in which generalised uncertainty relations are obtained without modified commutation relations. 
While existing models introduce modified phase space volumes for the canonical degrees of freedom, we introduce new degrees of freedom for the background geometry. 
The phase space is therefore enlarged but remains Euclidean. 
The spatial background is treated as a genuinely quantum object, with an associated state vector, and the model naturally gives rise to the extended generalised uncertainty principle (EGUP). 
Importantly, this approach solves (or rather, evades) well known problems associated with modified commutators, including violation of the equivalence principle, the `soccer ball' problem for multiparticle states, and the velocity dependence of the minimum length. 
However, it implies two radical conclusions. 
The first is that space must be quantised on a different scale to matter and the second is that the fundamental quanta of geometry are fermions.
We explain how, in the context of the model, these do not contradict established results including the no go theorems for multiple quantisation constants, which still hold for species of material particles, and the spin-$2$ nature of gravitons. 
\end{abstract}

\section*{}
{\it To appear in Touring the Planck Scale: Antonio Aurilia Memorial Volume, P. Nicolini, ed., Fundamental Theories of Physics, Springer}

\newpage

\tableofcontents 

\newpage

\section{Introduction} \label{sec:1}
 
The existence of geometric superpositions is the key requirement of any consistent quantum gravity theory \cite{DeWittMorette:2011zz,Lake:2018zeg}. 
It emerges as a logical necessity by combining the principles of general relativity (GR) with the principles of quantum mechanics (QM). 
In the former, gravity is described as the curvature of spacetime, which is sourced by the presence of matter \cite{Hobson:2006se}. 
In the latter, material particles exist in superpositions of position states \cite{DiracQM:book}. 
Therefore, if quantum matter is to act as the source of the gravitational field, spacetime geometries must also exist in superposition, and gravity must be quantised \cite{Lake:2018zeg,Marletto:2017pjr}. 

In addition, a widely expected feature of quantum gravity is the existence of a minimum length scale, of the order of the Planck length $l_{\rm Pl} \simeq 10^{-33}$ cm \cite{Hossenfelder:2012jw}. 
Unlike the existence of geometric superpositions, this is not a logical necessity of any viable theory. 
Nonetheless, general phenomenological arguments suggest that combining the essential features of both GR and QM limits the resolvability of physical measurements to super-Planck scales \cite{Garay:1994en,Padmanabhan:1985jq}. 
Therefore, phenomenological approaches to quantum gravity implement a minimum resolvable length scale, but do not link this to an underlying formalism describing superpositions of geometries \cite{Garay:1994en,Hossenfelder:2012jw}.

Many phenomenological arguments that give rise to a minimum length also suggest modifications of canonical QM and, in particular, of the canonical Heisenberg uncertainty principle (HUP). 
This forbids simultaneous knowledge of both position and momentum to arbitrary precision, such that
\begin{eqnarray} \label{HUP-1}
\Delta x^{i} \, \Delta p_j \gtrsim \frac{\hbar}{2} \delta^{i}{}_{j} \, ,
\end{eqnarray}  
where $\hbar = 1.054 \times 10^{-34} \, {\rm J \, . \, s}$ is the reduced Planck's constant. 
In the canonical theory, $\hbar$ represents the fundamental scale of action at which quantum effects become significant for material systems \cite{Rae}. 

The HUP can be introduced, heuristically, using the Heisenberg microscope thought experiment,
in which the irremovable uncertainty is explained as the result of momentum transferred to a massive particle by a probing photon \cite{aHeisenberg:1927zz,Heisenberg:1930,Rae}. 
More rigorously, it can be derived from the Hilbert space structure of the quantum state space, as shown by the pioneering work of Robertson and Schr\"odinger \cite{Robertson:1929zz,Schrodinger1930,Schrodinger:1930ty}.
In the latter, it is seen to arise from the general inequality
\begin{eqnarray} \label{Robertson-Schrodinger-1}
\Delta_\psi O_1 \, \Delta_\psi O_2 \geq \frac{1}{2} |\braket{\psi | [\hat{O}_1,\hat{O}_2] | \psi}|  \, ,
\end{eqnarray}  
where the uncertainty of the observable $\hat{O}$ is defined as the standard deviation $\Delta_\psi O = \sqrt{ \braket{\psi | \hat{O}^2 | \psi} - \braket{\psi | \hat{O} | \psi}^2 }$, plus the canonical commutator between position and momentum,
\begin{eqnarray} \label{[x,p]-1}
[\hat{x}^i,\hat{p}_j] = i\hbar\delta^{i}{}_{j} \ \hat{\mathbb{I}} \, .
\end{eqnarray}
A more careful statement of the HUP, derived from the underlying formalism of canonical QM, therefore reads
\begin{eqnarray} \label{HUP-2}
\Delta_\psi x^i \, \Delta_\psi p_j  \geq 
\frac{\hbar}{2} \delta^{i}{}_{j} \, ,
\end{eqnarray}
where $\Delta_\psi x^i$ and $\Delta_\psi p_j$ represent well defined standard deviations, unlike the heuristic uncertainties $\Delta x^i$ and $\Delta p_j$ in Eq. (\ref{HUP-1}).
 
Equation (\ref{HUP-2}) is more fundamental than (\ref{HUP-1}) because it holds in general, for any physical measurement scheme \cite{Ish95}. 
Nonetheless, as the example of Heisenberg's microscope shows, thought experiments can be useful in guiding our intuition. 
More recently, the microscope argument has been generalised to include the gravitational interaction between the massive particle and the photon, giving rise to the so called generalised uncertainty principle (GUP), 
\begin{eqnarray} \label{GUP-1}
\Delta x^i \gtrsim \frac{\hbar}{2\Delta p_j} \delta^{i}{}_{j} \left[1 + \alpha_0 \frac{2G}{\hbar c^3}(\Delta p_j)^2\right] \, ,
\end{eqnarray} 
where $\alpha_0$ is a numerical constant of order unity \cite{Adler:1999bu,Maggiore:1993rv,Scardigli:1999jh}. 
The GUP (\ref{GUP-1}) implies the existence of a minimum position uncertainty, of the order of the Planck length, but introduces a fundamental asymmetry between position and momentum \cite{Tawfik:2014zca,Tawfik:2015rva}. 
This is in stark contrast to the canonical HUP, which treats position and momentum on an equal footing. 

The microscope argument can also be generalised to account for the effects of a nonzero vacuum energy. 
In this scenario, the thought experiment is repeated in the presence of an asymptotically de Sitter space background \cite{Spradlin:2001pw}, in which the minimum scalar curvature is of the order of the cosmological constant, $\Lambda \simeq 10^{-56} \, {\rm cm^{-2}}$ \cite{Hobson:2006se}. 
We recall that this gives rise to a minimum vacuum energy density, $\rho_{\Lambda} = \Lambda c^2/(8\pi G) \simeq 10^{-30} \, {\rm g \, . \, cm^{-3}}$ \cite{Aghanim:2018eyx,Betoule:2014frx,Perlmutter1999,Reiss1998}. 
In \cite{Bambi:2007ty,Bolen:2004sq,Park:2007az}, it was argued that the presence of dark energy, in the form of a cosmological constant, modifies the canonical momentum uncertainty such that 
\begin{eqnarray} \label{EUP-1}
\Delta p_j \gtrsim \frac{\hbar}{2\Delta x^i} \delta^{i}{}_{j} \left[1 + 2\eta_0 \Lambda (\Delta x^i)^2\right] \, , 
\end{eqnarray} 
where $\eta_0$ is of order one. 
This relation, known in the literature as the extended uncertainty principle (EUP) \cite{Bambi:2007ty,Park:2007az}, implies a minimum momentum uncertainty of the order $\sim \hbar\sqrt{\Lambda}$. 
Thus, taking the GUP (\ref{GUP-1}) and EUP (\ref{EUP-1}) together reintroduces position-momentum symmetry into the gravitationally modified uncertainty relations. 

Furthermore, generalising the microscope argument to include the effects of both canonical gravity and repulsive dark energy implies the so called extended generalised uncertainty principle (EGUP) \cite{Bambi:2007ty,Park:2007az},
\begin{eqnarray} \label{EGUP-1}
\Delta x^i\Delta p_j \gtrsim \frac{\hbar}{2} \delta^{i}{}_{j} \left[1 + \alpha_0 \frac{2G}{\hbar c^3}(\Delta p_j)^2 + 2\eta_0\Lambda (\Delta x^i)^2\right] \, . 
\end{eqnarray} 
This implements both a minimum length and a minimum momentum in nature which corresponds to the existence of two, distinct, fundamental scales \cite{Lake:2018zeg}. 
The former is of order $l_{\rm Pl} \simeq 10^{-33} \, {\rm cm}$ whereas the latter is of the order of the de Sitter momentum, $m_{\rm dS}c \simeq 10^{-56} \, {\rm g \, . \, cm \, s^{-1}}$, which represents the minimum possible momentum uncertainty of a particle confined within the de Sitter horizon of the universe \cite{Spradlin:2001pw}. 

Considering these results, it may be hoped that progress towards a fundamental theory of quantum gravity can be made by replacing the heuristic uncertainties in Eqs. (\ref{GUP-1})-(\ref{EGUP-1}) with well defined standard deviations, derived from a non-canonical quantum theory. 
This forms the basic motivation for the subfield of quantum gravity research that deals with generalised uncertainty relations (GURs) \cite{Tawfik:2014zca,Tawfik:2015rva}. 

For more than 25 years, this field of research has been very active. 
To date, over 1200 papers on GURs have been published, with more than 50 appearing per year (on average), even today. 
As may be verified by a quick search on INSPIRE, the overwhelming majority of these studies utilise the same basic approach to the problem. 
In this, GURs are obtained directly by modifying the canonical commutation relations between position and momentum, (\ref{[x,p]-1}). 
This leads immediately to modifications of the Schr{\" o}dinger-Robertson bound for $\Delta_\psi x^i\Delta_\psi p_j$, and, hence, to modifications of the HUP (\ref{HUP-2}). 
Throughout, it is assumed that all measurement outcomes are derived from an underlying probability distribution, $|\psi|^2$, which depends only on the canonical degrees of freedom, $\left\{x^i\right\}_{i=1}^{3}$ or $\left\{p_j\right\}_{j=1}^{3}$. 
In other words, the operators and wavefunctions of modified commutator models are non-canonical, but the theories still describe material systems, without reference to any form of quantum geometry. 
This is at odds with their basic motivation, in which the non-canonical terms in the GUP, EUP and EGUP, (\ref{GUP-1})-(\ref{EGUP-1}), are expected to arise as the result of interactions between canonical quantum particles and the degrees of freedom of the background \cite{Adler:1999bu,Bambi:2007ty,Bolen:2004sq,Garay:1994en,Hossenfelder:2012jw,Maggiore:1993rv,Park:2007az,Scardigli:1999jh,Tawfik:2014zca,Tawfik:2015rva}. 

Moreover, even after more than a quarter of a century of effort, models based on modified commutation relations remain beset by conceptual and theoretical difficulties. 
These include their implied violation of the equivalence principle (EP) \cite{Tawfik:2014zca,Tawfik:2015rva}, and of Poincar{\' e} invariance in the relativistic limit \cite{Hossenfelder:2012jw}, the inability to construct a sensible multiparticle limit, otherwise known as the `soccer ball' problem \cite{Amelino-Camelia:2014gga,Gnatenko2019,Hossenfelder:2014ifa,LakeUkraine2019}, and the fact that even the most rigorous GUR models (see, for example, \cite{Kempf:1994su}) cannot reproduce, exactly, the GUP, EUP or EGUP obtained from gedanken experiment arguments.
\footnote{The latter point is subtle, but important, and is discussed in detail in Sec. \ref{subsubsec:2.1.3}.}  

In this work, we introduce a new formalism that, tentatively, offers solutions to each of the problems outlined above. 
To achieve this, we go back to basics, and seek to insert `at the ground level' the fundamental ingredients missing from existing phenomenological models. 
The most fundamental of these ingredients are the new quantum degrees of freedom, corresponding to the background space, which allow for the possibility of geometric superpositions \cite{Lake:2018zeg,Lake:2019nmn}. 

The structure of this paper is as follows. 
In Sec. \ref{sec:2}, we give a detailed account of the problems associated with existing modified commutator models. 
These include issues that have been discussed at length in the literature, which we review only briefly in Secs. \ref{subsubsec:2.1.1}-\ref{subsubsec:2.1.2}, as well as less well explored topics, which are considered in Secs. \ref{subsubsec:2.1.3}-\ref{subsubsec:2.1.4}. 
This provides our motivation for the new formalism, which is introduced in Sec. \ref{sec:3}. 
The basic conceptual and mathematical architecture of the model is given in Sec. \ref{sec:3.1}, and generalised position measurements are defined, leading to our derivation of the GUP. 
Generalised momentum measurements are defined in Sec. \ref{sec:3.2}, yielding our derivation of the EUP. 
The EGUP is derived, utilising both generalised position and momentum measurements, in Sec. \ref{sec:3.3}, while Sec. \ref{sec:3.4} considers the implications of our model for the measurement problem of canonical QM \cite{Ish95}. 
Sec. \ref{sec:4.1} is devoted to a unitarily equivalent formalism, which allows us to define generalised operators for angular momentum and spin, giving rise to additional GURs for these observables, respectively, in Secs. \ref{sec:4.2}-\ref{sec:4.3}. 
The relation of our model to the theory of quantum reference frames (QRFs) \cite{Giacomini:2017zju} is discussed in Sec. \ref{sec:4.4}. 
A brief summary of our results, including the current status of the model as well as prospects for future work, is given in Sec. \ref{sec:5}. 

Throughout the text, we use lower case letters to denote observables in canonical QM, for example $\hat{x}^i$ and $\hat{p}_j$.
Capital letters are used to denote generalised operators that give rise to GURs. 
Hence, $\hat{X}^i$ and $\hat{P}_j$ are used to denote position and momentum operators that obey modified commutation relations \cite{Hossenfelder:2012jw,Kempf:1993bq,Kempf:1994su,Tawfik:2014zca,Tawfik:2015rva} and to denote the alternative position and momentum operators proposed in the new theory \cite{Lake:2018zeg,Lake:2019nmn}. 
The different approaches are discussed in separate sections, so that the precise meaning of these symbols should be clear from the context. 
Where necessary, definitions are explicitly restated, to avoid confusion. 
The Planck length and mass scales are defined as 
\begin{eqnarray} \label{Planck_scales}
l_{\rm Pl} = \sqrt{\frac{\hbar G}{c^3}} \simeq 10^{-33} \, {\rm cm} \, , \quad m_{\rm Pl} = \sqrt{\frac{\hbar c}{G}} \simeq 10^{-5} \, {\rm g} \, ,
\end{eqnarray}
and we define the de Sitter length and mass scales as 
\begin{eqnarray} \label{dS_scales}
l_{\rm dS} = \sqrt{\frac{3}{\Lambda}} \simeq 10^{28} \, {\rm cm} \, , \quad m_{\rm dS} = \frac{\hbar}{c}\sqrt{\frac{\Lambda}{3}} \simeq 10^{-66} \, {\rm g} \, ,
\end{eqnarray}
for later convenience.   

\section{Problems with existing models - the need for a new approach} \label{sec:2}

Before outlining the new approach to GURs we discuss its motivation, namely, the outstanding theoretical problems faced by existing models based on modified commutation relations. 

\subsection{Problems with modified commutators} \label{subsec:2.1}

In this section, we consider four problems associated with the modified commutator approach to GURs. 
The first two, violation of the EP and the soccer ball problem for multiparticle states, have been extensively discussed in the existing literature \cite{Amelino-Camelia:2014gga,Gnatenko2019,Hossenfelder:2014ifa,LakeUkraine2019,Tawfik:2014zca,Tawfik:2015rva}. 
We review them only briefly in Secs. \ref{subsubsec:2.1.1}-\ref{subsubsec:2.1.2}. 
The third, concerning the violation of Galilean boost invariance and the consequent velocity dependence of the `minimum' length, has been touched upon in the literature (see, for example, \cite{Girdhar:2020kfl}), but not systematically explored. 
In Sec. \ref{subsubsec:2.1.3}, we give a detailed treatment of this problem and determine the exact form of the boost-dependent position uncertainty. 
The final problem is subtle and, to the best of our knowledge, has not been discussed previously. 
It concerns the supposedly `quantum' nature of the nonlocal modifications of the phase space geometry that give rise to GURs \cite{Nicolini:2010dj,Spallucci:2006zj}. 
This is discussed in Sec. \ref{subsubsec:2.1.4}

\subsubsection{Violation of the equivalence principle} \label{subsubsec:2.1.1}

In canonical QM, the Heisenberg equation for the time evolution of a hermitian operator $\hat{O}$ is 
\begin{eqnarray} \label{Heisenberg_Eq}
\frac{d}{dt}\hat{O}(t) = \frac{i}{\hbar}[\hat{H},\hat{O}] + \left(\frac{\partial\hat{O}}{\partial t}\right)_{H} \, ,
\end{eqnarray}
where $\hat{H} = \hat{p}^2/(2m) + V(\hat{\bf{x}})$. 
For the position operator $\hat{x}^{i}(t)$, this gives 
\begin{eqnarray} \label{x(t)}
\frac{d}{dt}\hat{x}^{i}(t) = \frac{\hat{p}^{i}}{m} \, .
\end{eqnarray}
The right-hand side follows directly from the identity $[AB,C] = A[B,C] + [A,C]B$, plus the form of the canonical position-momentum commutator (\ref{[x,p]-1}).

From (\ref{x(t)}), it follows that the acceleration of the position expectation value of a quantum particle is independent of its mass, i.e.,
\begin{eqnarray} \label{}
\frac{\hat{f}^{i}}{m} = \frac{1}{m}\frac{d\hat{p}^{i}}{dt} = \frac{d^2\hat{x}^{i}}{dt^2} \, .
\end{eqnarray}

For generalised operators $\hat{X}^{i}$ and $\hat{P}_{j}$, satisfying the modified commutation relation
\begin{eqnarray} \label{[X,P]-1}
[\hat{X}^i,\hat{P}_j] = i\hbar\delta^{i}{}_{j}G(\bf{\hat{\bf{P}}}) \, , 
\end{eqnarray}
the Heisenberg equation for $\hat{X}^{i}(t)$ is 
\begin{eqnarray} \label{X(t)}
\frac{d}{dt}\hat{X}^{i}(t) = \frac{\hat{P}^{i}}{m}G(\hat{\bf{P}}) \, .
\end{eqnarray}
Hence, for $G(\hat{\bf{P}}) \neq 1$, the particle experiences a mass-dependent acceleration,
\begin{eqnarray} \label{F^i}
\frac{\hat{F}^{i}}{m} = \frac{1}{m}\frac{d\hat{P}^{i}}{dt} = \frac{1}{G(\hat{\bf{P}})}\left[\frac{d^2\hat{X}^{i}}{dt^2} - \frac{\hat{P}^{i}}{m}\frac{dG(\hat{\bf{P}})}{dt}\right] \, ,
\end{eqnarray}
in clear violation of the EP. 
This conclusion assumes that the generalised Hamiltonian takes the form $\hat{H} = \hat{P}^2/(2m) + V(\hat{\bf{X}})$ and that a well defined Heisenberg picture exists in the generalised theory, but these are certainly reasonable assumptions \cite{Tawfik:2014zca,Tawfik:2015rva}. 
Similar analyses demonstrate that the EP is violated in modified commutator models defined by the functions $G(\hat{\bf{X}})$ and $G(\hat{\bf{X}},\hat{\bf{P}})$. 
It is therefore impossible to obtain the GUP, EUP, or EGUP from modified commutation relations without violating the founding principles of classical gravity \cite{Hobson:2006se}.   

\subsubsection{The soccer ball problem} \label{subsubsec:2.1.2}

The macroscopic limit of canonical QM obeys the correspondence principle \cite{Rae}. 
For single particle states, and with the exception of observables related to spin, two arbitrary hermitian operators $\hat{O}_1$ and $\hat{O}_2$ may be written as functions of the canonical variables $\left\{\hat{x}^{i},\hat{p}_{j}\right\}_{i,j=1}^{3}$. 
These are required to obey the correspondence limit,
\begin{eqnarray} \label{correspondence-1}
\lim_{\hbar \rightarrow 0}\frac{1}{i\hbar}[\hat{O}_1,\hat{O}_2] = \left\{O_1,O_2\right\}_{\rm PB} \, .
\end{eqnarray}
Here, $O_1$ and $O_2$ are have analogous functional dependence on the classical phase space coordinates, $\left\{x^{i},p_{j}\right\}_{i,j=1}^{3}$, and PB denotes the classical Poisson bracket. 
The general correspondence limit therefore requires that 
\begin{eqnarray} \label{correspondence-2}
\lim_{\hbar \rightarrow 0}\frac{1}{i\hbar}[\hat{x}^i,\hat{p}_j] = \left\{x^i,p_j\right\}_{\rm PB} = \delta^{i}{}_{j} \, , 
\end{eqnarray}
which forms the basis of the canonical quantisation procedure \cite{DiracQM:book}.

From Eq. (\ref{correspondence-1}), it is clear that the correspondence limit for modified commutator models implies analogous modifications of the classical Poisson brackets \cite{Tawfik:2014zca,Tawfik:2015rva}. 
This implies the violation of Galilean invariance, even for macroscopic systems, and, therefore, the violation of Poincar{\' e} invariance in the relativistic limit. 
Furthermore, modifications of the canonical position-momentum commutator can be expressed in terms of nonlinear corrections to the canonical de Broglie relation, ${\bf p}({\bf k})$ \cite{Hossenfelder:2012jw,Hossenfelder:2014ifa}. 
The wave number operator, $\hat{{\bf k}} = \hat{{\rm d}}_{{\bf x}}$, is the global shift-isometry generator for Euclidean position space \cite{Ish95}, and shift-isometries form a subgroup of the Galilean symmetry group \cite{Frankel:1997ec,Nakahara:2003nw}.   
Thus, in the relativistic regime, it is unclear whether one should require the physical momentum ${\bf p}$, or wave number ${\bf k}$, also known as the pseudo-momentum, to transform under the Poincar{\' e} group. 
In either case, the Lorentz transformations become nonlinear functions of the relevant quantity \cite{Hossenfelder:2012jw,Hossenfelder:2014ifa}. 

Arguably, it is more reasonable to choose ${\bf p}$ as the Lorentz invariant quantity. 
In this case, the nonlinear composition function is chosen to have a maximum at $\sim m_{\rm Pl}c$, which corresponds to the existence of a minimum length of order $l_{\rm Pl}$ \cite{Hossenfelder:2012jw,Hossenfelder:2014ifa}. 
This prevents single particles from ever exceeding the Planck momentum and so prevents the position uncertainties of their wave packets from becoming smaller than the Planck length. 
Unfortunately, since the sum of momenta can never exceed the maximum value, this also prevents multiparticle states from possessing momenta in excess of $m_{\rm Pl}c$.
\footnote{It should be noted that \cite{Amelino-Camelia:2014gga} offers an interesting counter-argument to this claim, at least within the framework of spacetime non-commutativity. However, to the best of our knowledge, these arguments have not been extended to models with arbitrary deformations of the momentum space.}

It is therefore unclear whether mutliparticle states with macroscopic momenta can be constructed in models based on modified commutation relations, and the problem of reproducing a sensible multiparticle limit is known as the soccer ball problem \cite{Amelino-Camelia:2014gga,Gnatenko2019,Hossenfelder:2014ifa,LakeUkraine2019}. 
Ultimately, this stems from the breaking of Galilean boost invariance in the non-relativistic regime. 
In the next section, we consider another problem raised by the violation of Galilean boost invariance, which also has important implications for modified commutator models. 

\subsubsection{Velocity-dependent uncertainties} \label{subsubsec:2.1.3}

In this section we show how non-relativisitc velocity boosts in one of the most commonly used modified commutator models, the GUP model proposed by Kempf, Mangano and Mann (KMM) \cite{Kempf:1994su}, give rise to velocity-dependent position uncertainties. 
This leads, automatically, to a reference frame-dependent minimum length. 
For clarity, we first review the standard treatment of velocity boosts and spatial translations in canonical QM, before generalising to the KMM model. 

In canonical QM the generators of translations in position and momentum space are defined by their actions on the Hilbert space bases $\ket{\bf x}$ and $\ket{\bf p}$, respectively:
\begin{eqnarray} \label{QM_translation_ops-1}
U({\bf x'})\ket{{\bf x}} = \ket{{\bf x - x'}} \, , \quad \tilde{U}({\bf p'}) \ket{{\bf p}} = \ket{{\bf p - p'}} \, .
\end{eqnarray}
They may be written explicitly as \cite{Ish95}
\begin{eqnarray} \label{QM_translation_ops-2}
U({\bf x'}) = \exp\left(-\frac{i}{\hbar} {\bf x'}.\hat{{\bf p}}\right) \, , \quad \tilde{U}({\bf p'}) = \exp\left(-\frac{i}{\hbar} \hat{{\bf x}}.{\bf p'}\right) \, , 
\end{eqnarray}
where 
\begin{eqnarray} \label{x_p}
\hat{{\bf x}} = \int {\bf x} \ket{{\bf x}}\bra{{\bf x}} {\rm d}^3{\rm x} \, , \quad \hat{{\bf p}} = \int {\bf p} \ket{{\bf p}}\bra{{\bf p}} {\rm d}^3{\rm p} 
\end{eqnarray}
are the vector generalisations of the canonical position and momentum operators, i.e., 
\begin{eqnarray} \label{QM_vector_ops}
\hat{{\bf x}} = \hat{x}^i \, {\bf e}_i(0) \, , \quad \hat{{\bf p}} = \hat{p}_j \, {\bf e}^j(0) \, . 
\end{eqnarray}
Here, $\left\{{\bf e}_i(x) \right\}_{i=1}^3$ $\left(\left\{{\bf e}^j(x) \right\}_{j=1}^3\right)$ denote the set of tangent (cotangent) vectors at the point $x$ in (classical) physical space \cite{Frankel:1997ec,Nakahara:2003nw}. 
The position space coordinates $\left\{x^i \right\}_{i=1}^3$ are taken to be Cartesians, so that $\left\{p_j \right\}_{j=1}^3$ represent the projections of the physical momentum ${\bf p}$ onto Cartesian axes in real space. 
The Hilbert space bases obey the `normalisation' conditions
\begin{eqnarray}
\braket{{\bf x}|{\bf x'}} = \delta^{3}({\bf x} - {\bf x'}) \, , \quad \braket{{\bf p}|{\bf p'}} = \delta^{3}({\bf p} - {\bf p'}) \, ,
\end{eqnarray}
which give rise to the canonical resolutions of the identity 
\begin{eqnarray} \label{res_ident}
\int \ket{{\bf x}}\bra{{\bf x}} {\rm d}^3{\rm x} = \int \ket{{\bf p}}\bra{{\bf p}} {\rm d}^3{\rm p} = \hat{\mathbb{I}} \, .
\end{eqnarray}

From here on we focus on the action of $\tilde{U}({\bf p'})$ since a translation in momentum space is equivalent to a Galilean velocity boost, up to a factor of $m$, where $m$ the mass of the system. 
When boosting by ${\bf v'} ={\bf p'}/m$, it is straightforward to show that the operator $\hat{{\bf p}}^n$ ($n \in \mathbb{Z}$) transforms as
\begin{eqnarray} \label{p_op_transf}
\hat{{\bf p}}^n \mapsto \tilde{U}({\bf p'}) \hat{{\bf p}}^n \tilde{U}^{\dagger}({\bf p'}) = (\hat{{\bf p}} + {\bf p'})^n \, . 
\end{eqnarray}
Equivalently, we may say that the state vector transforms as $\ket{\psi} \mapsto \tilde{U}^{\dagger}({\bf p'}) \ket{\psi}$ while $\hat{{\bf p}}$ remains unchanged \cite{Jones98}, but we adopt Eq. (\ref{p_op_transf}) for convenience. 
It follows immediately that
\begin{eqnarray}
\braket{\hat{{\bf p}}}_{\psi} \mapsto \braket{\tilde{U}({\bf p'}) \hat{{\bf p}} \tilde{U}^{\dagger}({\bf p'})}_{\psi} = \braket{\hat{{\bf p}}}_{\psi} + {\bf p'} \, , 
\nonumber\\
\braket{\hat{{\bf p}}^2}_{\psi} \mapsto \braket{\tilde{U}({\bf p'}) \hat{{\bf p}}^2 \tilde{U}^{\dagger}({\bf p'})}_{\psi} = \braket{\hat{{\bf p}}^2}_{\psi} + 2{\bf p'}\braket{\hat{{\bf p}}}_{\psi} + {\bf p'}^2 \, , 
\end{eqnarray}
so that the momentum uncertainty $\Delta_{\psi}p = \sqrt{\braket{\hat{{\bf p}}^2}_{\psi} - \braket{\hat{{\bf p}}}_{\psi}^2}$ is invariant. 
Similar arguments demonstrate the invariance of $\Delta_{\psi}x$, as well as the invariance of both uncertainties under translations in position space \cite{Ish95}. 
By projecting $(\Delta_{\psi}p)^2$ and $(\Delta_{\psi}x)^2$ onto the Cartesian axes, the invariance of the individual components $\Delta_{\psi}p_j$ and $\Delta_{\psi}x^i$ can also be demonstrated. 

In the KMM model the modified commutator takes the form
\begin{eqnarray} \label{KMM_mod_comm}
[\hat{X}^i, \hat{P}_j] = i\hbar \delta^{i}{}_{j}(1+\alpha {\bf \hat{P}}^2) \hat{\mathbb{I}} \, ,
\end{eqnarray}
where $\alpha = \alpha_0(m_{\rm Pl}c)^{-2}$ and $\alpha_0$ is a dimensionless constant of order one \cite{Kempf:1994su}. 
The generalised vector operators, 
\begin{eqnarray} \label{KMM_vector_ops}
\hat{{\bf X}} = \hat{X}^i {\bf e}_i(0) \, , \quad \hat{{\bf P}} = \hat{P}_j {\bf e}^j(0) \, , 
\end{eqnarray}
are defined by analogy with their counterparts in the canonical theory. 
This gives rise to the modified uncertainty relation
\begin{eqnarray} \label{KMM_GUP}
\Delta_{\psi}X^i \Delta_{\psi}P_j \geq \frac{\hbar}{2}\delta^{i}{}_{j}(1 + \alpha[(\Delta_{\psi}{\bf P})^2 + \braket{\hat{{\bf P}}}_{\psi}^2]) \, . 
\end{eqnarray}

Equation (\ref{KMM_GUP}) is almost the same as the GUP predicted by model-independent arguments, Eq. (\ref{GUP-1}), but, as we will now demonstrate, the presence of a term proportional to $\braket{\hat{{\bf P}}}_{\psi}^2$ on the right-hand side is crucial. 
For a given value of $\braket{\hat{{\bf P}}}_{\psi}^2$ the KMM GUP yields a minimum value of the position uncertainty and a corresponding critical value of the momentum uncertainty \cite{Kempf:1994su}:
\begin{eqnarray} \label{KMM_min_length}
(\Delta_{\psi}X^i)_{\rm min} = \hbar\sqrt{\alpha(1+\alpha \braket{\hat{{\bf P}}}_{\psi}^2)} \, , \quad (\Delta_{\psi}P_j)_{\rm crit} = 1/\sqrt{\alpha(1+\alpha \braket{\hat{{\bf P}}}_{\psi}^2)} \, .
\end{eqnarray}
For momentum-symmetric states, i.e., those for which $\braket{\hat{{\bf P}}}_{\psi} = 0$, these values reduce to $(\Delta_{\psi}X^i)_{\rm min} = \hbar\sqrt{\alpha} \simeq l_{\rm Pl}$ and $(\Delta_{\psi}P_j)_{\rm crit} = 1/\sqrt{\alpha} \simeq m_{\rm Pl}c$, respectively. 
However, even if a state $\ket{\psi}$ is symmetric in one particular frame, for example, the lab frame of a quantum experiment, it will be asymmetric in all others. 
We now consider this issue in detail and determine the exact dependence of $\Delta_{\psi}X^i$ on the velocity of the observer, relative to the centre of mass of the state $\ket{\psi}$.

In \cite{Kempf:1994su} it was shown that the modified commutator (\ref{KMM_mod_comm}) is obtained by introducing a modified momentum space volume, $(1+\alpha {\bf P}^2)^{-1}{\rm d}^3{\rm P}$. 
This corresponds to a modified normalisation condition and a modified resolution of the identity, viz.:
\begin{eqnarray} \label{KMM_res_ident}
\braket{{\bf P}|{\bf P'}} = (1+\alpha {\bf P}^2)\delta^{3}({\bf P} - {\bf P'}) \, , \quad \int \ket{{\bf P}}\bra{{\bf P}} \frac{{\rm d}^3{\rm P}}{(1+\alpha{\bf P}^2)} = \hat{\mathbb{I}} \, .
\end{eqnarray}
In this formulation of the GUP the position space representation is not well defined so that no spectral representation of $\hat{\bf X}$ exists \cite{Kempf:1994su}. 
We are therefore unable to construct a direct analogue of the velocity boost generator, $\tilde{U}({\bf p'})$, as given in Eq. (\ref{QM_translation_ops-2}). 
Nonetheless, we may define the unitary operator that generates generalised momentum space translations, $\tilde{\mathcal{U}}({\bf P'})$, in a purely abstract manner, via its actions on the modified kets $\ket{{\bf P}}$. 
This is by analogy with Eq. (\ref{QM_translation_ops-1}). 
It is straightforward to verify that the required action is 
\begin{eqnarray} \label{KMM_p-space_transl}
\tilde{\mathcal{U}}({\bf P'}) \ket{{\bf P}} = \frac{(1+\alpha {\bf P}^2)^{1/2}}{(1+\alpha ({\bf P - P'})^2)^{1/2}}\ket{{\bf P - P'}} \, , 
\end{eqnarray}
This preserves the relations (\ref{KMM_res_ident}) which ensures that the unitarity condition holds, $\tilde{\mathcal{U}}({\bf P'})\tilde{\mathcal{U}}^{\dagger}({\bf P'}) = \tilde{\mathcal{U}}^{\dagger}({\bf P'}) \tilde{\mathcal{U}}({\bf P'}) = \hat{\mathbb{I}}$.

It is then straightforward to show that
\begin{eqnarray} \label{P_op_transf}
\hat{{\bf P}}^n \mapsto \tilde{\mathcal{U}}({\bf P'}) \hat{{\bf P}}^n \tilde{\mathcal{U}}^{\dagger}({\bf P'}) = (\hat{{\bf P}} + {\bf P'})^n \, .
\end{eqnarray}
The generalised momentum uncertainty, $\Delta_{\psi}{\bf P} = \sqrt{\braket{\hat{{\bf P}}^2}_{\psi} - \braket{\hat{{\bf P}}}_{\psi}^2}$, is therefore invariant under the generalised momentum space `translations' defined by Eq. (\ref{KMM_p-space_transl}). 
These represent non-relativistic velocity boosts in the KMM theory and are the generalisations of the Galilean velocity boosts defined in Eq. (\ref{QM_translation_ops-1}). 
The modified commutator (\ref{KMM_GUP}) then transforms as
\begin{eqnarray} \label{KMM_comm_transf}
[\hat{X}^i, \hat{P}_j] \mapsto \tilde{\mathcal{U}}({\bf P'})[\hat{X}^i, \hat{P}_j]\tilde{\mathcal{U}}^{\dagger}({\bf P'}) = i\hbar \delta^{i}{}_{j}(1+\alpha ({\bf \hat{P}}+{\bf P'})^2) \hat{\mathbb{I}} \, ,
\end{eqnarray}
which leads to ${\bf P'}$-dependence of the corresponding uncertainty principle. 
Since $\Delta_{\psi}P_j$ is invariant it is clear that this is due to the ${\bf P'}$-dependence of $\Delta_{\psi}X^i$. 

Let us denote the boosted position uncertainty as $\Delta_{\psi}X'^i({\bf P'})$ so that $\Delta_{\psi}X'^i(0) = \Delta_{\psi}X^i$, where $ \Delta_{\psi}X^i$ is the position uncertainty appearing in the standard expression, Eq. (\ref{KMM_GUP}). 
We then have
\begin{eqnarray} \label{}
\Delta_{\psi}X'^i({\bf P'}) \geq \frac{\hbar}{2 \Delta_{\psi}P_j}\delta^{i}{}_{j}(1 + \alpha[(\Delta_{\psi}{\bf P})^2 + (\braket{\hat{{\bf P}}}_{\psi} + {\bf P'})^2]) \, . 
\end{eqnarray}
Even if $\ket{\psi}$ is symmetric in the original frame of observation, that is, if the initial lab frame is chosen to coincide with the motion of the centre of mass of the system, the minimum position uncertainty seen by an observer moving with the relative velocity ${\bf V'} = {\bf P'}/m$ is
\begin{eqnarray} \label{}
(\Delta_{\psi}X'^i)_{\rm min}({\bf P'}) \simeq \hbar\sqrt{\alpha}\left(1 + \frac{\alpha {\bf P'}^2}{2}\right) \, . 
\end{eqnarray}
For $|{\bf P'}| \ll 1/\sqrt{\alpha} \simeq m_{\rm Pl}c$ the boost-dependent term is, of course, very small. Nonetheless, its presence clearly violates the Galilean boost invariance that emerges as the low velocity limit of Lorentz invariance \cite{Frankel:1997ec,Nakahara:2003nw}. 
It is therefore at odds with the founding principles of both special and general relativity \cite{SRFrench,Hobson:2006se}, even for one-particle states. 

Though it is possible that these symmetries may be broken due to quantum effects on the geometry of spacetime \cite{Hossenfelder:2012jw,Tawfik:2014zca,Tawfik:2015rva} we note that there is, intrinsically, nothing quantum mechanical about the physical space background of the KMM model. 
The geometry remains classical but its symmetries are unknown, as is the exact form of the metric, $g_{ij}(X)$, to which they correspond. 
The symmetries of the momentum space geometry are also unknown, as is the symplectic structure of the classical phase space corresponding to the modified Jacobian, $J = (1+\alpha {\bf P}^2)^{-1}$ \cite{Frankel:1997ec,Nakahara:2003nw}. 
Despite this, it is often claimed that modifications of the canonical commutators and phase space volumes correspond to `universal' corrections induced by quantum gravity. 
In Sec. \ref{subsubsec:2.1.4}, we give a critical examination of these claims and argue against this interpretation of the nonlocal phase space geometry.

\subsubsection{The geometry is not quantum} \label{subsubsec:2.1.4}

In the current literature, there are many references to the `quantum' geometry obtained by introducing modified phase space volumes. 
This motivates a host of nonlocal gravity models that, it is claimed, follow directly from the quantum gravity corrections implied by GURs. 
In this section, we offer a critical examination of the link between modified commutators and nonlocal geometry, and find that this claim must be qualified.

It has been observed that modified momentum space volumes may be obtained by acting with nonlocal operators on the position space representations of the canonical QM eigenfunctions, 
$\braket{{\bf x}|{\bf x}'} = \delta^{3}({\bf x-x'})$ and $\braket{{\bf x}|{\bf p}} = (2\pi\hbar)^{-3/2}e^{i{\bf p}.{\bf x}/\hbar}$. 
An oft used example is the operator $e^{\sigma^2\Delta}$, where $\sigma$ is a fundamental length scale and $\Delta$ is the Laplacian \cite{Nicolini:2010dj,Nicolini:2012eu,Spallucci:2006zj,Sprenger:2012uc}. 
For convenience, we rewrite this in the spectral representation as
\begin{eqnarray} \label{nonlocal_op}
\hat{\zeta} = e^{-\hat{H}_0\Delta t/\hbar} \, , 
\end{eqnarray}
where $\hat{H}_0 = |\hat{{\bf p}}|^2/2m$ is the canonical free particle Hamiltonian, with $\hat{{\bf p}}$ given by Eq. (\ref{x_p}), and 
\begin{eqnarray} \label{spectral_rep_nonlocal_op}
\Delta t = \frac{2m\sigma^2}{\hbar} \, . 
\end{eqnarray}
The operator $\hat{\zeta}$ reduces to $e^{\sigma^2\Delta}$ in the wave mechanics picture but we may use Eq. (\ref{spectral_rep_nonlocal_op}) to define its action directly on the basis vectors $\ket{{\bf x}}$ and $\ket{{\bf p}}$. 
This, in turn, can be used to define a set of generalised basis vectors, $\ket{{\bf X}}$ and $\ket{{\bf P}}$, such that
\begin{eqnarray} \label{<X|X'>}
&&\braket{{\bf x}|\hat{\zeta}|{\bf x}'} = e^{\sigma^2\Delta}\braket{{\bf x}|{\bf x}'} = \left(\frac{1}{\sqrt{2\pi} \sigma}\right)^3e^{-({\bf x-x'})^2/2\sigma^2}  
\nonumber\\
&\equiv& \braket{{\bf X}|{\bf X}'} = \left(\frac{1}{\sqrt{2\pi} \sigma}\right)^3e^{-({\bf X-X'})^2/2\sigma^2} \, , 
\end{eqnarray}
and
\begin{eqnarray} \label{}
&&\braket{{\bf x}|\hat{\zeta}|{\bf p}} = e^{\sigma^2\Delta}\braket{{\bf x}|{\bf p}} = \left(\frac{1}{\sqrt{2\pi\hbar}}\right)^3 e^{-{\bf p}^2/2\tilde{\sigma}^2} e^{i{\bf p}.{\bf x}/\hbar} 
\nonumber\\
&\equiv& \braket{{\bf X}|{\bf P}} = \left(\frac{1}{\sqrt{2\pi\hbar}}\right)^3 e^{-{\bf P}^2/2\tilde{\sigma}^2} e^{i{\bf P}.{\bf X}/\hbar} \, ,
\end{eqnarray}
where $\tilde{\sigma} = \hbar/\sqrt{2}\sigma$. 
Consistency then requires the $\ket{{\bf P}}$ kets to satisfy a modified normalisation condition and a modified resolution of the identity \cite{Nicolini:2012eu,Sprenger:2012uc},
\begin{eqnarray} \label{}
\braket{{\bf P} | {\bf P'}} = e^{-{\bf P}^2/2\tilde{\sigma}^2}\delta^{3}({\bf P-P'}) \, , \quad \int \ket{{\bf P}}\bra{{\bf P}} e^{{\bf P}^2/2\tilde{\sigma}^2} {\rm d}^3{\rm P} = \hat{\mathbb{I}} \, . 
\end{eqnarray}
These differ from the conditions of the KMM model but, as it also modifies the momentum space volume element, $\hat{\zeta} \equiv e^{\sigma^2\Delta}$ naturally gives rise to a modified commutator that is a function of ${\bf P}$ \cite{Nicolini:2012eu,Sprenger:2012uc}. 

However, if we are not sufficiently careful, Eq. (\ref{<X|X'>}) can be misleading. 
We must be careful to distinguish between two inequivalent interpretations of $\delta^3({\bf X - X'})$:

\begin{itemize}

\item As a representation of the canonical position eigenfunction, $\braket{{\bf X}|{\bf X'}} = \delta^3({\bf X-X'})$. 
This has dimensions of ${\rm (length)}^{-3}$ and can be written as a superposition of momentum eigenstates, 
$\delta^3({\bf X - X'}) = (2\pi\hbar)^{-3}\int e^{i{\bf P}.({\bf X-X'})/\hbar} {\rm d}^3{\rm P}$, where ${\bf P} = \hbar{\bf K}$.

\item As a representation of a classical point source, $\delta^3({\bf X-X'})$. 
This also has dimensions of ${\rm (length)}^{-3}$ and admits a formal decomposition into plane wave modes as $\delta^3({\bf X - X'}) = (2\pi)^{-3}\int e^{i{\bf K}.{\bf X}} {\rm d}^3{\rm K}$. 
In this case, $ (2\pi)^{-3}\int e^{i{\bf K}.{\bf X}} {\rm d}^3{\rm K} \neq (2\pi\hbar)^{-3}\int e^{i{\bf P}.({\bf X-X'})/\hbar} {\rm d}^3{\rm P}$, since ${\bf P} = \hbar{\bf K}$ is not applicable.

\end{itemize}
It is important to note that delta functions of the second type have nothing to do with quantum mechanics, but may appear as source terms in the Poisson equation of classical Newtonian gravity \cite{Hobson:2006se}. 

In the existing literature, it is claimed that the link between the GURs and nonlocal gravity is provided by the semi-classical approach \cite{Moller:1959bhz,Rosenfeld:1963}. 
Here, it is assumed that curvature is sourced by the expectation value of the energy-momentum tensor operator, $\braket{\psi | \hat{T}_{\mu\nu} | \psi}$, but that gravity is described by the classical Einstein tensor, $G_{\mu\nu}$, i.e.,
\begin{eqnarray} \label{semi-classical-1}
R_{\mu\nu} - \frac{1}{2} R \, g_{\mu\nu} = \frac{8\pi G}{c^4} \braket{\psi | \hat{T}_{\mu\nu} | \psi} \, , 
\end{eqnarray}
where $R_{\mu\nu}$ is the Ricci tensor and $R$ is the scalar curvature. 
In the weak field limit, the semi-classical field equations (\ref{semi-classical-1}) reduce to \cite{Diosi:2014ura}
\begin{eqnarray} \label{semi-classical-2}
\nabla^2 {\rm \Phi} = 4\pi G m |\psi |^2 \, . 
\end{eqnarray}
It is then noted that the zero-width limit of the wave function, $\Delta_{\psi}{\bf X} \rightarrow 0$, yields a delta function source term, i.e., that
\begin{eqnarray} \label{semi-classical-3}
\lim_{\Delta_{\psi}{\bf X} \rightarrow 0}|\psi |^2 = \delta^3({\bf X-X'}) \, .
 \end{eqnarray}
The standard procedure, therefore, is to substitute $|\psi |^2 = \delta^3({\bf X-X'})$ into Eq. (\ref{semi-classical-2}) and, interpreting the source term as a position eigenfunction, to act on this with the nonlocal operator $e^{\sigma^2\Delta}$. 
On this basis, it is claimed that GURs imply nonlocal gravity and, furthermore, that the latter arises from `quantum' corrections to the classical theory \cite{Nicolini:2010dj,Nicolini:2012eu,Spallucci:2006zj,Sprenger:2012uc}. 

However, after substituting $|\psi |^2 = \delta^3({\bf X-X'})$ on the right-hand side, Eq. (\ref{semi-classical-2}) is equivalent to the Poisson equation for a classical point mass \cite{Hobson:2006se}. 
Acting on the point mass source term with $e^{\sigma^2\Delta}$ turns canonical Newtonian gravity into a classical nonlocal gravity theory, but it is important to recognise that no quantum corrections are implied. 
To obtain a true semi-classical theory of gravity, one must solve Eq. (\ref{semi-classical-2}) for states with nonzero width. 
This yields the self-interaction potential for the well known Schr{\" o}dinger-Newton equation in which ${\rm \Phi}$ becomes a function of $|\psi |^2$ \cite{Diosi:2014ura,Kelvin:2019esx}. 

It may easily be verified that, in the limit $\Delta_{\psi}{\bf X} \rightarrow 0$, the Schr{\" o}dinger-Newton potential reduces to the standard Newtonian potential generated by a classical point mass \cite{Diosi:2014ura,Kelvin:2019esx}. 
In this sense, Eq. (\ref{semi-classical-2}) remains semi-classical only for properly normalised states, with $\Delta_{\psi}{\bf X} > 0$. 
In principle, it is then possible to create a nonlocal theory of semi-classical gravity by acting with $\hat{\zeta} \equiv e^{\sigma^2\Delta}$ on states in the Schr{\" o}dinger-Newton Hilbert space. 
The latter step, however, is completely optional, and is in no way implied by the existence of classical nonlocal gravity theories. 
To the best of our knowledge, it has not been attempted.

The action of the nonlocal operator on a $\delta^3({\bf X-X'})$ source term therefore `blows up' a classical point mass into a classical sphere of finite density. 
If the blow up is sufficiently strong, the sphere acquires an effective equation of state that makes it stiff enough to resist gravitational collapse. 
In this way, nonlocal gravity models are able to cure the singularity problem encountered in canonical general relativity, although this also requires the standard energy conditions to be violated \cite{Nicolini:2012eu}. 
This part of the usual analysis is valid, but its connection to GURs must now be qualified. 

The considerations above show that nonlocal operators `blow up' sections of the classical phase space. 
This turns local classical theories into nonlocal classical theories, but does not incorporate quantum corrections in any way. 
Similarly, the action of nonlocal operators turns local quantum theory, which does not permit the superluminal transfer of information, into nonlocal quantum theory \cite{Nicolini:2012eu,Sprenger:2012uc}. 
The HUP exists in the former, whereas GURs are manifested in the latter via the existence of modified commutators. 
The claim that the GURs imply nonlocal gravity is therefore suspect. 
Instead, it is more accurate to claim that classical nonlocal gravity and modified commutators have the same underlying cause, viz. modifications of the classical phase space volumes, over which both classical densities and quantum mechanical amplitudes (wave functions) must be integrated. 

In both cases, the number of degrees of freedom remains the same as in the corresponding local theory. 
No new quantum degrees of freedom are introduced. 
In this sense, the delocalisation of classical phase space points affected by $e^{\sigma^2\Delta}$, and other nonlocal operators, is also manifestly classical. 
It affects both classical and quantum theories but is not, in itself, intrinsically quantum in nature.

\subsection{What the new approach must achieve} \label{subsec:2.2}

In short, the new approach must address all of the outstanding theoretical issues discussed in Secs. \ref{subsubsec:2.1.1}-\ref{subsubsec:2.1.4}. 
It must either solve, or evade, the soccer ball problem and must not give rise to mass-dependent accelerations, or to velocity-dependent uncertainties. 
Finally, the new model should generate GURs via the introduction of new, genuinely quantum, degrees of freedom. 
These should describe the quantum state of the background geometry as a vector in an appropriate Hilbert space. 
In Sec. \ref{sec:3}, we outline the basic formalism of a new model that, tentatively, provides solutions to each of these problems. 

\section{First formalism - position and linear momentum} \label{sec:3}

In this section, we outline the first formalism of the smeared space model, developed in \cite{Lake:2018zeg}. 
This treats matter and geometry as entangled and allows us to successfully derive GURs for both position and linear momentum, giving rise to the EGUP. 

\subsection{Generalised position measurements} \label{sec:3.1}

In \cite{Lake:2018zeg}, a new model of quantum geometry was proposed in which each point ${\bf x}$ in the classical background is associated with a vector $\ket{g_{{\bf x}}}$ in a Hilbert space, where  
\begin{eqnarray} \label{g_x}
\ket{g_{{\bf x}}} = \int g({\bf x}'-{\bf x}) \ket{{\bf x}'} {\rm d}^{3}{\rm x}' \, , 
\end{eqnarray}
and $g(\bf{x}'-\bf{x})$ is any normalised function, i.e.,  
\begin{eqnarray} \label{g_normalisation}
\int |g({\bf x}'-{\bf x})|^2 {\rm d}^{3}{\rm x} = \int |g({\bf x}'-{\bf x})|^2 {\rm d}^{3}{\rm x}' = 1 \, . 
\end{eqnarray}

The motivation for this identification is simple. 
As discussed in Sec. \ref{subsubsec:2.1.4}, classical nonlocal geometries may be generated by first identifying each point ${\bf x'}$ with a Dirac delta, $\delta^3({\bf x - x'})$.
Roughly speaking, this is equivalent to stating that the point ${\bf x'}$ is where it should be, in relation to all other points ${\bf x}$, with 100\% certainty. 
Classical nonlocality is then introduced by smearing each delta into a finite-width probability distribution, $P({\bf x-x'})$. 
(For example, a normalised Gaussian, as in Eq. (\ref{<X|X'>}).)
Note that, in this case, ${\bf x'}$ is simply a parameter that determines the position of the distribution, whereas ${\bf x}$ is its genuine argument. 
No new degrees of freedom are added.

Thus, in order to introduce a genuinely quantum form of nonlocality, we instead associate the point ${\bf x'}$ with the rigged basis vector of a Hilbert space, $\ket{{\bf x'}}$. 
The latter is then `smeared' to produce the normalised state $\ket{g_{{\bf x}}}$, given by Eq. (\ref{g_x}). 
Importantly, $\braket{{\bf x'} | g_{{\bf x}}} = g({\bf x}'-{\bf x})$ is a genuine quantum mechanical amplitude, not a probability distribution. 
It has dimensions of ${\rm (length)}^{-3/2}$ rather than ${\rm (length)}^{-3}$ and, potentially, contains nontrivial phase information. 
This is the first crucial difference between the smeared space model and the standard nonlocal geometry theories. 

However, it is not enough. 
In \cite{Lake:2018zeg}, it was shown that mapping $\ket{{\bf x}} \mapsto \ket{g_{{\bf x}}}$, alone, cannot generate valid probabilities. 
The problem is that, in canonical QM, $\ket{{\bf x}'}$ represents the state of quantum particle which is ideally localised at the classical point ${\bf x'}$. 
It does not represent the quantum state of a spatial point per se. 
To move forward, we must introduce new degrees of freedom that explicitly associate quantum state vectors with points in the classical space. 
By integrating over the states of individual points, we may then associate a quantum state with the geometry as a whole. 
We therefore begin by associating every classical point ${\bf x}$ with a basis vector $\ket{{\bf x}}$. 
This represents the ideally localised state of a `point' in the quantum geometry, in the position space representation. 
We then smear each ideally localised point into a superposition of all points via the map
\begin{eqnarray} \label{smearing_map}
S: \ket{{\bf x}} \mapsto \ket{{\bf x}} \otimes \ket{g_{{\bf x}}} \, .
\end{eqnarray}

We may visualise the smearing map (\ref{smearing_map}) as follows: for each point ${\bf x} \in \mathbb{R}^3$ in the classical geometry we obtain one whole `copy' of $\mathbb{R}^3$, doubling the size of the classical phase space. 
The resulting smeared geometry is represented by a six-dimensional volume, $\mathbb{R}^3 \times \mathbb{R}^3$, in which each point $({\bf x},{\bf x}')$ is associated with a quantum probability amplitude, $g({\bf x}'-{\bf x})$. 
This is interpreted as the amplitude for the transition ${\bf x} \leftrightarrow {\bf x}'$ and the higher-dimensional space is interpreted as a superposition of three-dimensional geometries \cite{Lake:2018zeg,Lake:2019nmn}. 
The correspondence between the classical phase space and quantum phase space of the model is
\begin{eqnarray} \label{}
{\bf x} \leftrightarrow \ket{{\bf x}} \, , \quad {\rm d}^3{\bf x} \leftrightarrow \ket{{\bf x}}{\rm d}^3{\bf x} \, , \quad ( \, . \, , \, . \, ) \leftrightarrow . \, \otimes \, . \, ,
\end{eqnarray}
where the pairing $( \, . \, , \, . \, )$ represents the Cartesian product of two manifolds \cite{Lake:2018zeg}. 
In other words, we introduce additional degrees of freedom at the level of the classical phase space by allowing $\mathbb{R}^3 \mapsto \mathbb{R}^3 \times \mathbb{R}^3$, then map Cartesian products to tensor products in a natural way.  

Each geometry in the smeared superposition of geometries is flat, but differs from all others by the pair-wise exchange of two points \cite{Lake:2018zeg,Lake:2019nmn}.
\footnote{In the original smeared space formalism, presented in \cite{Lake:2018zeg}, more general transitions of the form ${\bf x} \rightarrow {\bf x}'$ were considered. These permit metric fluctuations with nonzero curvature, which are expected to arise in the complete theory of quantum gravity \cite{Crowell:2005ax}. However, the inclusion of these fluctuations is inconsistent with the leading-order approximation of GR in the non-relativistic regime, in which the Newtonian gravitational potential is treated as a scalar field on a flat Euclidean background \cite{Hobson:2006se}. For this reason, we restrict our attention to transitions of the form ${\bf x} \leftrightarrow {\bf x}'$ in the weak field limit, which is consistent with the existence of flat space. See \cite{Lake:2019nmn} for further discussion of this point.} 
In other words, it is assumed that the interchange of points ${\bf x} \leftrightarrow {\bf x}'$ exchanges the associated canonical amplitudes, $\psi({\bf x}) \leftrightarrow \psi({\bf x}')$, but that this leaves the curvature of the space unchanged. 
In this limit, the back-reaction generated by the presence of canonical quantum matter, described by the wave function $\psi$, is neglected. 
This is consistent with the weak field limit of classical gravity, in which the gravitational potential is treated, formally, as a scalar potential on a flat Euclidean background  \cite{Hobson:2006se}. 
However, in this model, `points' in the spatial background exist in a superposition of states, and may undergo stochastic fluctuations as the result of measurements. 
This affects the statistics of the canonical quantum matter living on, or `in', the space, including the behaviour of quantum particles under the influence of different potentials \cite{Lake:2018zeg}.
\footnote{The interested reader is referred to \cite{Lake:2018zeg,Lake:2019nmn} for full details of the smearing procedure. The procedure for smearing an arbitrary potential in canonical QM, and for obtaining the corresponding generalised Schr{\" o}dinger equation for the composite matter-plus-geometry state vector, is outlined in \cite{Lake:2018zeg}.} 

For simplicity, we may imagine $|g({\bf x}'-{\bf x})|^2$ as a normalised Gaussian centred on ${\bf x}' = {\bf x}$, but, here, ${\bf x}'$ is no longer just a parameter. 
By introducing the tensor product structure (\ref{smearing_map}), we have doubled the number of degrees of freedom, vis-{\` a}-vis canonical QM. 
Those in the left-hand subspace are labelled by ${\bf x}$, whereas those in the right-hand subspace are labelled by ${\bf x'}$. 
This is the second crucial difference between the smeared space model and the standard nonlocal geometry theories. 

The interaction between the primed and unprimed degrees of freedom describes the interaction between canonical quantum matter and the quantum background geometry in which it propagates. 
Thus, the action of $S$ on $\ket{{\bf x}}$ (\ref{smearing_map}) induces a map on the canonical quantum state vector, $\ket{\psi} = \int \psi({\bf x})\ket{{\bf x}} {\rm d}^{3}{\rm x}$, such that
\begin{eqnarray} \label{psi->Psi}
S : \ket{\psi} \mapsto \ket{\Psi} \, , 
\end{eqnarray}
where
\begin{eqnarray} \label{|Psi>_position_space}
\ket{\Psi} = \int\int \psi({\bf x}) g({\bf x}'-{\bf x}) \ket{{\bf x},{\bf x}'} {\rm d}^{3}{\rm x}{\rm d}^{3}{\rm x}' \, ,
\end{eqnarray}
and $\ket{{\bf x},{\bf x}'} = \ket{{\bf x}} \otimes \ket{{\bf x}'}$. 
The square of the smeared-state wave function $\Psi({\bf x},{\bf x}') = \braket{{\bf x},{\bf x}'|\Psi}$, i.e., $|\Psi({\bf x},{\bf x}')|^2 = |\psi({\bf x})|^2|g({\bf x}'-{\bf x})|^2$, represents the probability distribution associated with a quantum particle propagating in the quantum geometry. 
Specifically, since $|\psi({\bf x})|^2$ represents the probability of finding the particle at the fixed classical point ${\bf x}$ in canonical QM, $|\psi({\bf x})|^2|g({\bf x}'-{\bf x})|^2$ represents the probability that it will now be found, instead, at a new point ${\bf x}'$. 
If $g$ is a Gaussian centred on the origin, ${\bf x}'={\bf x}$ remains the most likely value, but fluctuations within a volume of order $\sim \sigma_g^{3}$, where $\sigma_g$ is the standard deviation of 
$|g|^2$, remain relatively likely \cite{Lake:2018zeg}. 

In this model, delocalised spatial points exhibit `wave-point duality' and their relative positions are subject to stochastic fluctuations, induced by coherent transitions between ideally localised states \cite{Lake:2018zeg,Lake:2019nmn}. 
We now show that, since $g({\bf x}'-{\bf x})$ must have finite width, $\sigma_g > 0$, according to the normalisation condition (\ref{g_normalisation}), this naturally implements a minimum resolvable length scale. 

Since an observed value ` ${\bf x}'$ ' cannot determine which point(s) underwent the transition ${\bf x} \leftrightarrow {\bf x}'$ in the smeared superposition of geometries, we must sum over all possibilities by integrating the joint probability distribution $|\Psi({\bf x},{\bf x}')|^2 = |\psi({\bf x})|^2 |g({\bf x}'-{\bf x})|^2$ over ${\rm d}^3{\rm x}$, yielding 
\begin{eqnarray} \label{EQ_XPRIMEDENSITY}
\frac{{\rm d}^{3}P({\bf x}' | \Psi)}{{\rm d}{\rm x}'^{3}} = \int |\Psi({\bf x},{\bf x}')|^2 {\rm d}^3{\rm x} = (|\psi|^2 * |g|^2)({\bf x}') \, ,
\end{eqnarray}
where the star denotes a convolution. 
Here, physical predictions are assumed to be those of the smeared space theory and the canonical QM of the original (unprimed) degrees of freedom is only a convenient tool in our calculations. 
In this formalism, only primed degrees of freedom represent measurable quantities, whereas unprimed degrees of freedom are physically inaccessible \cite{Lake:2018zeg}.

The variance of a convolution is equal to the sum of the variances of the individual functions, so that the probability distribution (\ref{EQ_XPRIMEDENSITY}) gives rise to an uncertainty relation that is not of the canonical Heisenberg type. 
It is straightforward to verify that the same statistics can be obtained from the generalised position-measurement operator $\hat{X}^{i}$, defined as
\begin{eqnarray} \label{X_operator}
\hat{X}^{i} = \int x'^{i} \, {\rm d}^{3} \hat{\mathcal{P}}_{\vec{x}'} = \hat{\mathbb{I}} \otimes \hat{x}'^{i} \, ,
\end{eqnarray}
where ${\rm d}^{3}\hat{\mathcal{P}}_{\vec{x}'} = \hat{\mathbb{I}} \otimes \ket{{\bf x}'}\bra{{\bf x}'}{\rm d}^{3}{\rm x}'$.
We then have
\begin{eqnarray} \label{X_uncertainty}
(\Delta_\Psi X^{i})^2 = \braket{\Psi |(\hat{X}^{i})^{2}|\Psi} - \braket{\Psi|\hat{X}^{i}|\Psi}^2 = (\Delta_\psi x'^{i})^2 + (\Delta_gx'^i)^2 \, .
\end{eqnarray}
Next, we note that the HUP, expressed here in terms of primed variables,
\begin{eqnarray} \label{HUP}
\Delta_{\psi} x'^{i} \Delta_{\psi} p'_{j} \geq \frac{\hbar}{2} \delta^{i}{}_{j} \, , 
\end{eqnarray}
(recall that the unprimed degrees of freedom are physically inaccessible), holds independently of Eq. (\ref{X_uncertainty}) \cite{Lake:2018zeg}. 
Substituting this into Eq. (\ref{X_uncertainty}), identifying the width of $|g|^2$ with the Planck length, 
\begin{equation} \label{sigma_g} 
\Delta_{g}x'^{i} = \sqrt{2}l_{\rm Pl} \, ,  
\end{equation}   
and Taylor expanding the resulting expression to first order, then yields
\begin{eqnarray} \label{smeared_GUP}
\Delta_\Psi X^{i} \gtrsim \frac{\hbar}{2\Delta_{\psi} p'_{j}}\delta^{i}{}_{j}\left[1 + \alpha(\Delta_{\psi} p'_{j})^2\right] \, , 
\end{eqnarray}
where $\alpha = 4(m_{\rm Pl}c)^{-2}$ \cite{Lake:2018zeg}. 
In this expression, $(\Delta_{\psi} p'_{j})^2 = (\Delta_{\psi} p'_{j}) \, . \, (\Delta_{\psi} p'^{j})$, but no sum is implied by the repeated index.

For $\Delta_{\psi} x'^{i} \gg \Delta_{g} x'^{i} \simeq l_{\rm Pl}$, we have that $\Delta_\Psi X^{i} \simeq \Delta_{\psi} x'^{i}$ (\ref{X_uncertainty}). 
In this limit, Eq. (\ref{smeared_GUP}) reduces to the GUP derived from gedanken experiment arguments (\ref{GUP-1}), but with the heuristic uncertainties $\Delta x^i$ and $\Delta p_j$ replaced by well defined standard deviations. 
Importantly, the smeared space GUP does not contain a term proportional to $\braket{\hat{{\bf P}}}_{\psi}^2$, unlike the KMM GUP (\ref{KMM_GUP}). 
It is therefore compatible with Galilean symmetry and does not lead to the kind of velocity-dependent uncertainties described in Sec. \ref{subsubsec:2.1.3}. 

\subsection{Generalised momentum measurements} \label{sec:3.2}

In the momentum space picture, the composite matter-plus-geometry state vector may be expanded as 
\begin{eqnarray} \label{Psi_p}
\ket{\Psi} = \int\int \psi_{\hbar}({\bf p})\tilde{g}_{\beta}({\bf p}'-{\bf p}) \ket{{\bf p} \, {\bf p}'} {\rm d}^{3}{\rm p}{\rm d}^{3}{\rm p}' \, ,
\end{eqnarray}
where 
\begin{eqnarray} \label{dB-1}
\tilde{\psi}_{\hbar}({\bf p}) = \left(\frac{1}{\sqrt{2\pi\hbar}}\right)^3 \int \psi({\bf x}) e^{-\frac{i}{\hbar}{\bf p}.{\bf x}}{\rm d}^{3}{\rm x} \, , 
\end{eqnarray}
as in canonical QM, and
\begin{eqnarray} \label{}
\tilde{g}_{\beta}({\bf p}'-{\bf p}) = \left(\frac{1}{\sqrt{2\pi\beta}}\right)^3 \int g({\bf x}'-{\bf x}) e^{-\frac{i}{\beta}({\bf p}'-{\bf p}).({\bf p}'-{\bf p})}{\rm d}^{3}{\rm x}' \, ,  
\end{eqnarray}
where $\beta \neq \hbar$ is a new action scale \cite{Lake:2018zeg,Lake:2019nmn}. 

The momentum space representation of the canonical quantum wave function, $\psi_{\hbar}({\bf p})$, is given by the weighted Fourier transform of the position space representation, $\psi({\bf x})$. 
The transformation is performed at the scale $\hbar$, which is equivalent to assuming the standard de Broglie relation, ${\bf p} = \hbar{\bf k}$, for the matter sector. 
Here, we use the subscript $\hbar$ to emphasise this point. 
By contrast, the momentum space representation of the smearing function, $\tilde{g}_{\beta}({\bf p}'-{\bf p})$, is given by the weighted Fourier transform of $g({\bf x}'-{\bf x})$, where the transformation is performed at the new scale $\beta$. 
This represents the quantisation scale for space (rather than matter) and must be fixed by physical considerations. 
In \cite{Lake:2018zeg}, it was shown that, in order to reproduce the observed vacuum energy density, $\rho_{\Lambda} = \Lambda c^2/(8\pi G) \simeq 10^{-30} \, {\rm g \, . \, cm^{-3}}$, where $\Lambda \simeq 10^{-56} \, {\rm cm^{-2}}$ is the cosmological constant \cite{Hobson:2006se}, $\beta$ must take the order of magnitude value
\begin{eqnarray} \label{beta_mag}
\beta = 2\hbar\sqrt{\frac{\rho_{\Lambda}}{\rho_{\rm Pl}}} \simeq \hbar \times 10^{-61} \, , 
\end{eqnarray}
where $\rho_{\rm Pl} \simeq 10^{93}  \, {\rm g \, . \, cm^{-3}}$ is the Planck density. 

The consistency of Eqs. (\ref{|Psi>_position_space}) and (\ref{Psi_p}) requires
\begin{eqnarray} \label{mod_dB}
\braket{{\bf x},{\bf x}' | {\bf p} \, {\bf p}'} = \left(\frac{1}{2\pi\sqrt{\hbar\beta}}\right)^3 e^{\frac{i}{\hbar}{\bf p}.{\bf x}} e^{\frac{i}{\beta}({\bf p}'-{\bf p}).({\bf x}'-{\bf x})} \, .
\end{eqnarray}
Hence, $\ket{{\bf p} \, {\bf p}'}$ represents an entangled state in the rigged basis of the `enlarged' Hilbert space, $\mathcal{H} \otimes \mathcal{H}$, where $\mathcal{H}$ is the Hilbert space of canonical QM.
\footnote{In fact the Hilbert space is not enlarged, since $\mathcal{H} \otimes \mathcal{H} \cong \mathcal{H}$. The isomorphism holds because $\mathcal{H}$ is the unique Hilbert space with countably infinite dimensions, i.e., the Hilbert space of canonical QM in any number of (physical) spatial dimensions \cite{HilbertSpaces}.} 
We emphasise this by not writing a comma between ${\bf p}$ and ${\bf p}'$, by contrast with $\ket{{\bf x},{\bf x}'} = \ket{{\bf x}} \otimes \ket{{\bf x}'}$. 
By complete analogy with the position space representation, $\tilde{g}_{\beta}({\bf p}'-{\bf p})$ is interpreted as the quantum probability amplitude for the transition ${\bf p} \leftrightarrow {\bf p}'$ in smeared momentum space \cite{Lake:2018zeg,Lake:2019nmn}. 
Analogous reasoning to that presented in Sec. \ref{sec:3.2} then gives
\begin{eqnarray} \label{EQ_PPRIMEDENSITY}
\frac{{\rm d}^{3}P({\bf p}' | \tilde{\Psi})}{{\rm d}{\rm p}'^{3}} = \int |\tilde{\Psi}({\bf p},{\bf p}')|^2 {\rm d}^{3}{\rm p} = (|\tilde{\psi}_{\hbar}|^2 * |\tilde{g}_{\beta}|^2)({\bf p}') \, ,
\end{eqnarray}
and 
\begin{eqnarray} \label{P_operator}
\hat{P}_{j} = \int p'_{j} \, {\rm d}^{3}\hat{\mathcal{P}}_{\vec{p}{\, '}} \, ,
\end{eqnarray}
where ${\rm d}^{3}\hat{\mathcal{P}}_{{\bf p}'} = \left(\int \ket{{\bf p} \, {\bf p}'}\bra{{\bf p} \, {\bf p}'} {\rm d}^{3}{\rm p}\right){\rm d}^{3}{\rm p}'$. 
It follows that
\begin{eqnarray} \label{P_uncertainty}
(\Delta_\Psi P_{j})^2 = \braket{\Psi |(\hat{P}_{i})^{2}|\Psi} - \braket{\Psi|\hat{P}_{j}|\Psi}^2 = (\Delta_\psi p'_{j})^2 + (\Delta_g p'_j)^2 \, .
\end{eqnarray}
The general properties of the Fourier transform \cite{pinsky2008introduction} also ensure that the `wave-point' uncertainty relation, 
\begin{eqnarray} \label{Beta_UP}
\Delta_{g} x'^{i} \Delta_{g} p'_{j} \geq \frac{\beta}{2} \delta^{i}{}_{j} \, , 
\end{eqnarray}
holds independently of Eqs. (\ref{X_uncertainty}) and (\ref{P_uncertainty}), and of the HUP (\ref{HUP}). 

For convenience, we denote the position uncertainty of the smearing function as $\Delta_{g}x'^{i} = \sigma_{g}^{i}$ when $|g|^2$ is chosen to be a Gaussian. 
Since Gaussians Fourier transform to Gaussians, choosing $|g|^2$ to be a normal distribution in the position space representation implies that $|\tilde{g}_{\beta}|^2$ is a normal distribution in momentum space. 
In this case, we denote the momentum uncertainty as $\Delta_{g}p'_{j} = \tilde\sigma_{gj}$. 
The inequality (\ref{Beta_UP}) is saturated for Gaussian distributions, yielding the definition of the new transformation scale $\beta$:
\begin{eqnarray} \label{beta}
\beta = (2/3)\sigma_{g}^{i}\tilde\sigma_{gi} \, . 
\end{eqnarray}

The HUP contains the essence of wave-particle duality or, rather, wave-point-particle duality, and is a fundamental consequence of the canonical de Broglie relation ${\bf p}= \hbar {\bf k}$. 
This, in turn, is equivalent to the relation (\ref{dB-1}), which holds for particles propagating on a fixed (classical) Euclidean background.  
By contrast, Eq. (\ref{Beta_UP}) represents the uncertainty relation for quantised spatial `points' (not point-particles {\it in} space). 
This follows directly from Eq. (\ref{mod_dB}), which is equivalent to the modified de Broglie relation
\begin{eqnarray} \label{mod_dB*}
{\bf p}' = \hbar{\bf k} + \beta({\bf k}'-{\bf k}) \, .
\end{eqnarray}
The new relation holds for particles propagating in the smeared-space background and the non-canonical term may be interpreted, heuristically, as an additional momentum `kick' induced by quantum fluctuations of the background geometry  \cite{Lake:2018zeg,Lake:2019nmn}. 

Substituting the HUP into Eq. (\ref{P_uncertainty}), identifying the width of $|\tilde{g}_{\beta}|^2$ with the de Sitter momentum,
\begin{equation} \label{sigma_g*}
\Delta_{g}p'_{j} = \frac{1}{2}m_{\rm dS}c \, ,
\end{equation} 
and Taylor expanding to first order yields
\begin{eqnarray} \label{smeared_EUP}
\Delta_\Psi P_{j} \gtrsim \frac{\hbar}{2\Delta_{\psi} x'^{i}}\delta^{i}{}_{j}\left[1 + \eta(\Delta_{\psi} x'^{i})^2\right] \, ,
\end{eqnarray}
where $\eta = (1/2)l_{\rm dS}^{-2}$  \cite{Lake:2018zeg,Lake:2019nmn}. 
Here, $(\Delta_{\psi} x'^{i})^2 = (\Delta_{\psi} x'^{i}) \, . \, (\Delta_{\psi} x'_{i})$, but no sum is implied by the repeated index. 
For $\Delta_{\psi} p'_{j} \gg \Delta_{g} p'_{j} \simeq m_{\rm dS}c$, we have that $\Delta_\Psi P_{j} \simeq \Delta_{\psi} p'_{j}$ (\ref{P_uncertainty}). 
In this limit, Eq. (\ref{smeared_EUP}) reduces to the EUP derived from gedanken experiment arguments (\ref{EUP-1}), but with the heuristic uncertainties $\Delta x^i$ and $\Delta p_j$ replaced by well defined standard deviations, as in the derivation of the smeared space GUP given in Sec. \ref{sec:3.1}. 

From our analysis of the KMM GUP, presented in Sec. \ref{subsubsec:2.1.3}, it is clear that an analogous EUP model may be constructed by introducing a modified position space volume element, $(1+\eta {\bf X}^2)^{-1}{\rm d}^3{\rm X}$. 
In this scenario, the momentum space representation is not well defined, but an EUP that is similar in form to (\ref{smeared_EUP}) may be generated. 
The main difference is that the KMM-inspired EUP would, necessarily, include a term proportional to $\braket{\hat{{\bf X}}}_{\psi}^2$. 
This would violate the Galilean translation invariance of physical space, leading to position-dependent momentum uncertainties, just as the KMM GUP leads to momentum-dependent position uncertainties. 
We note that this problematic term is absent from EUP generated by the smeared space model. 

\subsection{The EGUP} \label{sec:3.3}

In Secs. \ref{sec:3.1}-\ref{sec:3.2}, we showed how the smeared space model is able to generate both the GUP and the EUP using well defined position and momentum space representations, respectively.
We now show how the smeared space GURs can be combined to give the EGUP. 
This result is important because, in the modified commutator approach to the EGUP, neither well defined position space nor well defined momentum space representations are available. 
Instead, one must resort to a generalised Bargman-Fock representation \cite{Kempf:1993bq,Kempf:1994su,Kempf:1996ss}. 
However, first, we show how the model naturally avoids other problems associated with modified commutators, including those presented in Secs. \ref{subsubsec:2.1.1}-\ref{subsubsec:2.1.4}. 

Equations (\ref{X_uncertainty}), (\ref{HUP}) and (\ref{P_uncertainty}) 
can be combined to give
\begin{eqnarray} \label{GUR_X}
(\Delta_\Psi X^{i})^2 \, (\Delta_\Psi P_{j})^2 
&\ge& (\hbar/2)^2(\delta^{i}{}_{j})^2 + (\Delta_\psi x'^{i})^2 (\Delta_{g} p'_{j})^2 
\nonumber\\
&+& (\Delta_{g}x'^{i})^2\frac{(\hbar/2)^2}{(\Delta_\psi x'^{j})^2} + (\Delta_{g}x'^{i})^2 (\Delta_{g} p'_{j})^2  \, ,
\end{eqnarray}
plus an analogous relation containing only $(\Delta_{\psi}p'_{j})^2$.
Optimising the right-hand side of (\ref{GUR_X}) with respect to $\Delta_\psi x'^{i}$, and~its counterpart with respect to $\Delta_\psi p'_{j}$, yields
\begin{equation} \label{EQ_CAN_DX_OPT}
(\Delta_\psi x'^{i})_{\mathrm{opt}} = \sqrt{\frac{\hbar}{2} \frac{\Delta_{g} x'^{i}}{\Delta_{g} p'_{i}}} \, , \quad (\Delta_\psi p'_{j})_{\mathrm{opt}} = \sqrt{\frac{\hbar}{2} \frac{\Delta_{g} p'_{j}}{\Delta_{g} x'^{j}}} \, , 
\end{equation}
so that
\begin{eqnarray} \label{DXDP_opt}
\Delta_\Psi X^{i} \, \Delta_\Psi P_{j} & \ge & \frac{(\hbar + \beta)}{2} \, \delta^{i}{}_{j} \, .
\end{eqnarray}
The same result is readily obtained by noting that the commutator of the generalised position and momentum observables is
\begin{equation} \label{[X,P]}
[\hat{X}^{i},\hat{P}_{j}] = i(\hbar + \beta)\delta^{i}{}_{j} \, {\bf\hat{\mathbb{I}}} \, .
\end{equation}
Equation (\ref{DXDP_opt}) then follows directly from the Schr{\" o}dinger--Robertson relation (\ref{Robertson-Schrodinger-1}). 

The inequalities in all five uncertainty relations, (\ref{X_uncertainty}), (\ref{HUP}), (\ref{P_uncertainty}), (\ref{Beta_UP}) and (\ref{GUR_X}), are saturated when $|g|^2$ is chosen to be a Gaussian, for which we denote $\Delta_{g}x'^{i} = \sigma_{g}^{i}$ and $\Delta_{g}p'_{j} = \tilde\sigma_{gj}$, and~when $|\psi|^2$ is chosen to be a Gaussian with $\Delta_\psi x'^{i} = (\Delta_\psi x'^{i})_{\rm opt}(\sigma_{g}^{i},\tilde\sigma_{gi})$, $\Delta_\psi p'_{j} = (\Delta_\psi p'_{j})_{\rm opt}(\sigma_{g}^{j},\tilde\sigma_{gj})$ (\ref{EQ_CAN_DX_OPT}). 
This yields the absolute limit, $\Delta_\Psi X^{i} \, \Delta_\Psi P_{j} = (\hbar + \beta)/2 \, . \, \delta^{i}{}_{j}$. 
Thus, in general, the smeared space model gives rise to an uncertainty relation of the form $\Delta_\Psi X^{i} \, \Delta_\Psi P_{j} \geq \dots \geq (\hbar + \beta)/2 \, . \, \delta^{i}{}_{j}$. 
The term on the far right-hand side is the generalised Schr{\" o}dinger--Robertson bound, which is simply a small rescaling of the Schr{\" o}dinger--Robertson bound derived in canonical QM, such that $\hbar \rightarrow \hbar + \beta$. 
The terms in the middle give rise to GURs. 

This shows that the GUP and the EUP may be obtained without introducing modified commutation relations of the type considered in the existing literature \cite{Kempf:1994su,Tawfik:2014zca,Tawfik:2015rva}. 
We stress that these GURs may be generated by a completely different mathematical structure, which is compatible with the canonical shift-isometry algebra (now rescaled by $\hbar + \beta$), as in Eq. (\ref{[X,P]}). 
This simple fact allows us to circumvent virtually all the theoretical problems associated with previous models, including violation of the EP, Sec. \ref{subsubsec:2.1.1}, the soccer ball problem for multiparticle states, Sec. \ref{subsubsec:2.1.2}, and the velocity dependence of the minimum length, Sec. \ref{subsubsec:2.1.3}. 
The problems associated with modified commutators have not been solved. 
In the smeared space model, they do not arise in the first place. 

In addition, the smeared space GUP and EUP are generated by introducing new quantum degrees of freedom for the background geometry. 
This circumvents many of the problems discussed in Sec. \ref{subsubsec:2.1.4}  
and allows the model to implement both minimum length and momentum uncertainties in the presence of commuting position and momentum space coordinates, i.e., 
\begin{equation} \label{XX_PP_commutators}
[\hat{X}^{i},\hat{X}^{j}] = 0 \, , \quad [\hat{P}_{i},\hat{P}_{j}] = 0 \, .
\end{equation}
The corresponding uncertainty relations take the form $\Delta_\Psi X^{i}\Delta_\Psi X^{j} \gtrsim l_{\rm Pl}^2\delta^{ij} \geq 0$ and $\Delta_\Psi P_{i}\Delta_\Psi P_{j} \gtrsim m_{\rm dS}^2c^2\delta_{ij} \geq 0$, respectively, where the terms on the far right-hand sides represent the Schr{\" o}dinger--Robertson limits. 

We now have the tools we need to derive the EGUP. 
This step is also important because, in one respect, the GUP and the EUP derived in Secs. \ref{sec:3.1} and \ref{sec:3.2}, respectively, remain unsatisfactory. 
Specifically, we note that the right-hand side of the GUP (\ref{smeared_GUP}) is written entirely in terms of $\Delta_{\psi} p'_{j}$. 
Similarly, the right-hand side of the EUP (\ref{smeared_EUP}) is written entirely in terms of $\Delta_{\psi} x'^{i}$. 
However, in the smeared space model, neither $\Delta_{\psi} x'^{i}$ nor $\Delta_{\psi} p'_{j}$ are directly measurable, and only $\Delta_{\Psi} X^{i}$ and $\Delta_{\Psi} P_{j}$ are physical \cite{Lake:2018zeg,Lake:2019nmn}. 
It is therefore useful to express the smeared space GUR directly in terms of these quantities. 

Directly combining Eqs. (\ref{X_uncertainty}), (\ref{HUP}) and (\ref{P_uncertainty}), we obtain 
\begin{eqnarray} \label{smeared-spaceEGUP-1}
(\Delta_{\Psi} X^{i})^2 (\Delta_{\Psi} P_{j})^2 &\geq& (\hbar/2)^2(\delta^{i}{}_{j})^2 + (\Delta_{g}x'^{i})^2(\Delta_{\Psi} P_{j})^2 
\nonumber\\
&+& (\Delta_{\Psi} X^{i})^2(\Delta_{g} p'_{j})^2 - (\Delta_{g}x'^{i})^2(\Delta_{g} p'_{j})^2 \, .
\end{eqnarray}
Substituting for $\Delta_{g}x'^{i}$ and $\Delta_{g}p'_{j}$ from Eqs. (\ref{sigma_g}) and (\ref{sigma_g*}), taking the square root and expanding to first order, then ignoring the subdominant term of order $\sim l_{\rm Pl}m_{\rm dS}c$, yields
\begin{eqnarray} \label{smeared-spaceEGUP-2}
\Delta_{\Psi} X^{i} \Delta_{\Psi} P_{j} \gtrsim \frac{\hbar}{2}\delta^{i}{}_{j}\left[1 + \alpha(\Delta_{\Psi} P_{j})^2 + \eta(\Delta_{\Psi} X^{i})^2\right] \, ,
\end{eqnarray}
where 
\begin{eqnarray} \label{smeared-spaceEGUP-3}
\alpha = \frac{4G}{\hbar c^3} \, , \quad \eta = \frac{\Lambda}{6} \, .
\end{eqnarray}
Here, $(\Delta_{\Psi} P_{j})^2 = (\Delta_{\Psi} P_{j}) \, . \, (\Delta_{\Psi} P^{j})$ and $(\Delta_{\Psi} X^{i})^2 = (\Delta_{\Psi} X^{i}) \, . \, (\Delta_{\Psi} X_{i})$, but no sum is implied over either repeated index. 
Equation (\ref{smeared-spaceEGUP-2}) is analogous to the EGUP obtained by model-independent gedanken experiments, Eq. (\ref{EGUP-1}), but with the heuristic uncertainties replaced by the standard deviations of well defined observables. 
This is the primary achievement of the first formalism.

Interestingly, this form of the EGUP is saturated when the position and momentum uncertainties of the canonical quantum matter take the optimum values provided by Eqs. (\ref{sigma_g}), (\ref{sigma_g*}) and (\ref{EQ_CAN_DX_OPT}), i.e.,
\begin{eqnarray} \label{}
(\Delta_\psi x')_{\rm opt} = l_{\Lambda} \, , \quad (\Delta_\psi p')_{\rm opt} = \frac{1}{2}m_{\Lambda}c \, , 
\end{eqnarray}
where $l_{\Lambda} \simeq \sqrt{l_{\rm Pl}l_{\rm dS}} \simeq 0.1 \ {\rm mm}$ and $m_{\Lambda} \simeq \sqrt{m_{\rm Pl}m_{\rm dS}} \simeq 10^{-3} \ {\rm eV}$. 
This gives rise to a minimum energy density of order 
\begin{eqnarray} \label{min_energy_density}
\mathcal{E}_\psi \simeq \frac{3}{4\pi}\frac{(\Delta_\psi p')_{\rm opt}\ c}{(\Delta_\psi x')^3_{\rm opt}} \simeq \rho_{\Lambda}c^2 = \frac{\Lambda c^4}{8\pi G} \, , 
\end{eqnarray}
as required by current cosmological data  \cite{Aghanim:2018eyx,Betoule:2014frx,Perlmutter1999,Reiss1998}. 
In other words, the wave function corresponding to a space-filling `sea' of dark energy particles, each of mass $m_{\Lambda} \simeq 10^{-3} \ {\rm eV}$ and Compton radius $l_{\Lambda} \simeq 0.1 \ {\rm mm}$, would minimise the right-hand side of Eq. (\ref{smeared-spaceEGUP-2}). 
In this scenario, dark energy would remain approximately constant over large distances, but may appear granular on sub-millimetre scales \cite{Antoniou:2017mhs,Burikham:2015nma,Hashiba:2018hth,Krishak:2020opb,Lake:2017ync,Lake:2017uzd,Lake:2018zeg,Lake:2020chb,Perivolaropoulos:2016ucs}. 

\subsection{Implications for the measurement problem} \label{sec:3.4}

To conclude our treatment of the first formalism, we note that the smeared space model has important implications for the description of measurement in quantum mechanics. 
We now illustrate these by considering a generalised position measurement, in detail. 
Applying the generalised position operator $\hat{{\bf X}} = \hat{X}^{i}\bold{e}_{i}(0)$ to an arbitrary pre-measurement state $\ket{\Psi}$ returns a random measured value, ${\bf x}'$, and projects the state in the fixed background subspace of the tensor product onto
\begin{eqnarray} \label{X_measurement}
\ket{\psi_{{\bf x}'}} = \frac{1}{(|\psi|^2*|g|^2)({\bf x}')}\int \psi({\bf x}) g({\bf x}'-{\bf x}) \ket{{\bf x}} {\rm d}^3{\rm x} \, , 
\end{eqnarray}
with probability $(|\psi|^2*|g|^2)({\bf x}')$ \cite{Lake:2018zeg}.
The total state is then $\ket{\psi_{{\bf x}'}}\otimes \ket{{\bf x}'}$, which is non-normalisable and therefore unphysical. 
This is analogous to the action of the canonical position measurement operator on $\ket{\psi}$, which projects onto the unphysical state $\ket{{\bf x}}$ with probability $|\psi({\bf x})|^2$. 

However, in the smeared space formalism, we must reapply the fundamental `smearing' map (\ref{smearing_map}) to complete our description of the measurement process \cite{Lake:2018zeg,Lake:2019nmn}. 
The smeared measurement may therefore be split into two parts. 
In the first, an ideal projective measurement is performed on the second subspace of the tensor product, which corresponds to the observable position ${\bf x}'$. 
This yields the measured value of position, but, as in canonical QM, the resulting state is unphysical. 
Re-application of the map (\ref{smearing_map}) the re-smears the ideally localised point in the quantum background geometry, giving rise to a finite width for the post-measurement composite state $\ket{\Psi}$ \cite{Lake:2018zeg,Lake:2019nmn}. 
 
Hence, although the generalised position measurements, represented by the application of the map (\ref{smearing_map}) to the state (\ref{X_measurement}), yield precise measurement values, the post-measurement states are always physical, with well defined norms. 
Their position uncertainties, which may be determined by performing multiple measurements on ensembles of identically prepared systems, never fall below the fundamental smearing scale, $\Delta_{g}x'^{i} \simeq l_{\rm Pl}$. 
Analogous considerations hold for generalised momentum measurements, with the corresponding minimum uncertainty $\Delta_{g}p'_{j} \simeq m_{\rm dS}c$. 

In this section, we have presented only a brief overview of the first smeared space formalism. 
The interested reader is referred to \cite{Lake:2018zeg,Lake:2019nmn} for further details.

\section{Second formalism - angular momentum and spin} \label{sec:4}

In this section, we present a second, unitarily equivalent, formalism for the smeared space model, originally developed in \cite{Lake:2019nmn}. 
This treats the composite matter-plus-geometry system as a separable state, and enables us to extend our analysis to include angular momentum and spin.

\subsection{(Un-)entangling algebras} \label{sec:4.1}

For the generalised position and momentum measurements defined in Secs. \ref{sec:3.1}-\ref{sec:3.2}, we were able to exploit a simple property of convolutions to obtain the GURs (\ref{X_uncertainty}) and (\ref{P_uncertainty}), respectively. 
In each case, the total variance split into the sum of the variances associated with the canonical QM wave function, $\psi$, and the geometric part, $g$. 
However, for more complicated functions of $\hat{{\bf X}}$ and $\hat{{\bf P}}$, the presence of the entangled momentum space basis $\ket{{\bf p} \, {\bf p}'}$ (\ref{mod_dB}) prevents such a neat decomposition \cite{Lake:2019nmn}. 
For this reason, it is useful to introduce a unitary transformation that `symmetrises' the position and momentum space bases. 

In the symmetrised bases, the composite matter-plus-geometry state $\ket{\Psi}$ becomes separable and both $\hat{{\bf X}}$ and $\hat{{\bf P}}$ split into the sum of two terms. 
The first terms act nontrivially only on the first subspace of the tensor product, whose quantum properties are determined by $\hbar$, whereas the second terms act nontrivially only on the second subspace, whose quantum properties are determined by $\beta$.
The rescaled Heisenberg algebra (\ref{[X,P]})-(\ref{XX_PP_commutators}), which is equivalent to the translation isometry algebra of Euclidean space, weighted by the factor $(\hbar + \beta)$, then splits into the sum of two commuting `copies'. 
The first copy is a representation of the shift-isometry algebra weighted by $\hbar$, and is therefore equivalent to the Heisenberg algebra of canonical QM, whereas the second copy is weighted by $\beta$. 

Thus, by unentangling matter and geometry in the composite state $\ket{\Psi}$, we also `unentangle' the rescaled algebra (\ref{[X,P]})-(\ref{XX_PP_commutators}). 
We show that this arises from the combination of two subalgebras, one of which holds for the matter sector and one of which holds for the quantum state of the geometry. 
Roughly speaking, the former symmetries generate the physical momenta of canonical quantum particles, whereas the latter generate the physical momentum carried by the quantum state of the background. 
Since the primed and unprimed degrees of the freedom interact via the modified de Broglie relation (\ref{mod_dB*}), these subalgebras combine to form the rescaled algebra for the composite matter-plus-geometry system, Eqs. (\ref{[X,P]})-(\ref{XX_PP_commutators}). 

The symmetrised bases also allow us to decompose the generalised angular momentum algebra in a similar way. 
In this case, there exist nontrivial cross terms, i.e., terms that act nontrivially on both subspaces of the tensor product, and the subalgebra structure is more complicated.
The variance of the generalised angular momentum operator, $(\Delta_{\Psi}L_{i})^2$, then splits into the sum of a pure matter part and a pure geometry part, plus covariance terms involving both the matter and geometry sectors. 
However, this is sufficient for our purposes. 
The first of these terms represents the contribution to the total uncertainty given by the canonical QM degrees of freedom. 
The rest are non-canonical and arise directly as a result of the smearing map (\ref{smearing_map}). 
This allows us to construct GURs of the form $\Delta_{\Psi}L_{i} \, \Delta_{\Psi}L_{j} \geq {\rm (QM \ terms) + (quantum \ geometry \ corrections)}$ in Sec. \ref{sec:4.2}. 
By analogy, we construct GURs for generalised spin measurements in Sec. \ref{sec:4.3}. 
These represent the two main achievements of the second formalism. 

Let us begin by constructing the unitary operator
\begin{eqnarray} \label{U_beta}
\hat{U}_{\beta} = \exp\left[-\frac{i}{\beta}(\hat{\mathbb{I}} \otimes \hat{{\bf p}}').(\hat{{\bf x}} \otimes \hat{\mathbb{I}}) \right] \, , 
\end{eqnarray}
whose action on the original smeared space basis is
\begin{eqnarray} \label{U_beta_action_pos}
\hat{U}_{\beta} \ket{{\bf x},{\bf x}'} = \ket{{\bf x},{\bf x}' - {\bf x}} \, ,
\end{eqnarray}
\begin{eqnarray} \label{U_beta_action_mom}
\hat{U}_{\beta} \ket{{\bf p} \,{\bf p}'} = \ket{{\bf p},{\bf p}' - {\bf p}} \, .
\end{eqnarray}
Here, we again assume that $\hbar$ sets the quantisation scale for the degrees of freedom in the first subspace of the tensor product, while $\beta$ sets the quantisation scale for the degrees of freedom in the second subspace. 
Hence, $\beta^{-1}(\hat{\mathbb{I}} \otimes \hat{{\bf p}}')$ generates translations on the second vector of the basis $\ket{{\bf x},{\bf x}'}$, just as $\hbar^{-1}(\hat{{\bf p}} \otimes \hat{\mathbb{I}})$ generates translations on the first. 
This accounts for Eq. (\ref{U_beta_action_pos}). 
Equation (\ref{U_beta_action_mom}) then follows by combining Eqs. (\ref{U_beta})--(\ref{U_beta_action_pos}) with Eq. (\ref{mod_dB}).

Together, Eqs. (\ref{mod_dB}) and (\ref{U_beta})--(\ref{U_beta_action_pos}) imply
\begin{eqnarray} \label{non-canon*}
\braket{{\bf x},{\bf x}' | {\bf p} \, {\bf p}'} = \braket{{\bf x} | {\bf p}}_1\braket{{\bf x}' - {\bf x} | {\bf p}' - {\bf p}}_2 \, , 
\end{eqnarray}
where
\begin{eqnarray} \label{non-canon-3}
\braket{{\bf x} | {\bf p}}_1 = \left(\frac{1}{\sqrt{2\pi\hbar}}\right)^3 e^{\frac{i}{\hbar}{\bf p}.{\bf x}} \, ,
\end{eqnarray}
as in canonical QM, and
\begin{eqnarray} \label{non-canon-4}
\braket{{\bf x}' - {\bf x} | {\bf p}' - {\bf p}}_2 = \left(\frac{1}{\sqrt{2\pi\beta}}\right)^3 e^{\frac{i}{\beta}({\bf p}'-{\bf p}).({\bf x}'-{\bf x})} \, .
\end{eqnarray}
In Eqs. (\ref{non-canon*})--(\ref{non-canon-4}), we use the subscripts 1 and 2 to indicate which subspace of the tensor product state the brakets belong to. 
This is to avoid confusion since, in these expressions, the degrees of freedom in each subspace are no longer labelled exclusively by primed or unprimed variables, as they were previously. 
Nonetheless, they remain consistent with our convention that $\hbar$ sets the quantisation scale for degrees of freedom in the first subspace of the tensor product, while $\beta$ sets the quantisation scale for degrees of freedom in the second. 
We repeat that the former are associated with canonical quantum matter whereas the latter are associated with the quantum state of the background geometry. 

Using these results, we map the smeared space operators $\hat{X}^{i}$ and $\hat{P}_{j}$, and the smeared state $\ket{\Psi}$, according to
\begin{eqnarray} \label{X_unitary_equiv}
\hat{X}^{i} \mapsto \hat{U}_{\beta} \hat{X}^{i} \hat{U}_{\beta}^{\dagger}  
= \int\int x'^{i} \ket{{\bf x}}\bra{{\bf x}} \otimes \ket{{\bf x}' - {\bf x}}\bra{{\bf x}' - {\bf x}}{\rm d}^3{\rm x} {\rm d}^3{\rm x}' 
\end{eqnarray}
\begin{eqnarray} \label{P_unitary_equiv}
\hat{P}_{j} \mapsto \hat{U}_{\beta} \hat{P}_{j} \hat{U}_{\beta}^{\dagger}  
= \int\int p'_{j} \ket{{\bf p}}\bra{{\bf p}} \otimes \ket{{\bf p}' - {\bf p}}\bra{{\bf p}' - {\bf p}}{\rm d}^3{\rm p} {\rm d}^3{\rm p}' 
\end{eqnarray}
and
\begin{eqnarray} \label{Psi_unitary_equiv}
\ket{\Psi} \mapsto \hat{U}_{\beta}\ket{\Psi}
&=& \int\int g({\bf x}' - {\bf x})\psi({\bf x}) \ket{{\bf x},{\bf x}'-{\bf x}} {\rm d}^3{\rm x} {\rm d}^3{\rm x}'
\nonumber\\
&=& \int\int \tilde{g}_{\beta}({\bf p}' - {\bf p})\tilde{\psi}_{\hbar}({\bf p}) \ket{{\bf p},{\bf p}'-{\bf p}} {\rm d}^3{\rm p} {\rm d}^3{\rm p}' 
\nonumber\\
&=& \ket{\psi} \otimes \ket{g} \, . 
\end{eqnarray}
Note that Eq. (\ref{Psi_unitary_equiv}) implicitly defines the state $\ket{g}$, which is distinct from the state $\ket{g_{{\bf x}}}$ defined in Eq. (\ref{g_x}). 
Physically, $\ket{g_{{\bf x}}}$ represents the quantum state associated with the smeared `point' ${\bf x}$, whereas $\ket{g}$ represents the quantum state associated with whole background space. 
From here on, we use $\hat{X}^{i}$, $\hat{P}_{j}$ and $\ket{\Psi}$ to refer to the unitarily equivalent forms of the generalised position and momentum operators, (\ref{X_unitary_equiv}) and (\ref{P_unitary_equiv}), and smeared state (\ref{Psi_unitary_equiv}), respectively, unless stated otherwise. 

Next, we split each of the generalised operators (\ref{X_unitary_equiv}) and (\ref{P_unitary_equiv}) into the sum of two terms as
\begin{eqnarray} \label{ops_split_Q}
\hat{X}^{i} = \hat{Q}^{i} + \hat{Q}'^{i} = (\hat{q}^{i} \otimes \hat{\mathbb{I}}) + (\hat{\mathbb{I}} \otimes \hat{q}'^{i}) \, , 
\end{eqnarray}
\begin{eqnarray} \label{ops_split_Pi}
\hat{P}_{j} = \hat{\Pi}_{j} + \hat{\Pi}'_{j} = (\hat{\pi}_{j} \otimes \hat{\mathbb{I}}) + (\hat{\mathbb{I}} \otimes \hat{\pi}'_{j}) \, , 
\end{eqnarray}
where
\begin{eqnarray} \label{QQ'}
\hat{Q}^{i} &=& (\hat{q}^{i} \otimes \hat{\mathbb{I}}) = \int\int x^{i} \ket{{\bf x}}\bra{{\bf x}} \otimes \ket{{\bf x}' - {\bf x}}\bra{{\bf x}' - {\bf x}}{\rm d}^3{\rm x} {\rm d}^3{\rm x}'  \, , 
\nonumber\\
\hat{Q}'^{i} &=& (\hat{\mathbb{I}} \otimes \hat{q}'^{i}) = \int\int (x'^{i} -x^{i}) \ket{{\bf x}}\bra{{\bf x}} \otimes \ket{{\bf x}' - {\bf x}}\bra{{\bf x}' - {\bf x}}{\rm d}^3{\rm x} {\rm d}^3{\rm x}' \, ,
\end{eqnarray}
and 
\begin{eqnarray} \label{PiPi'}
\hat{\Pi}_{j} &=& (\hat{\pi}_{j} \otimes \hat{\mathbb{I}})  = \int\int p_{j} \ket{{\bf p}}\bra{{\bf p}} \otimes \ket{{\bf p}' - {\bf p}}\bra{{\bf p}' - {\bf p}}{\rm d}^3{\rm p} {\rm d}^3{\rm p}' \, , 
\nonumber\\
\hat{\Pi}'_{j} &=& (\hat{\mathbb{I}} \otimes \hat{\pi}'_{j}) = \int\int (p'_{j} -p_{j}) \ket{{\bf p}}\bra{{\bf p}} \otimes \ket{{\bf p}' - {\bf p}}\bra{{\bf p}' - {\bf p}}{\rm d}^3{\rm p} {\rm d}^3{\rm p}' \, .
\end{eqnarray}
In other words, we define the new classical variables
\begin{eqnarray} \label{new_variables-1}
{\bf X} = {\bf x}' \, , \quad {\bf q} = {\bf x} \, , \quad {\bf q}' = ({\bf x}' - {\bf x}) \, , 
\end{eqnarray}
\begin{eqnarray} \label{new_variables-2}
{\bf P} = {\bf p}' \, , \quad \bm{\pi} = {\bf p} \, , \quad \bm{\pi}' = ({\bf p}' - {\bf p}) \, , 
\end{eqnarray}
and construct their quantum operator counterparts. 

Note that, in this formulation of the smeared space model, measurable quantities are no longer expressed in terms of primed variables only.  
That is, neither $q^{i}$ nor $q'^{i}$ are directly measurable, and only their sum $q^{i} + q'^{i} = x'^{i}$ is physical. 
Similarly, neither $\pi_{j}$ nor $\pi'_{j}$ is measurable individually, only $\pi_{j} + \pi'_{j} = p'_{j}$. 
This has important physical consequences. 

In the first formalism of the smeared space model \cite{Lake:2018zeg}, summarised in Sec. \ref{sec:3}, the wave functions corresponding to matter and geometry are entangled, as hypothesised in \cite{Kay:2018mxr}. 
However, in the alternative formalism presented here, they are not. 
Nonetheless, physical measurements are represented by operators that act on both subsytems of the tensor product state $\ket{\Psi}$, regardless of our choice of basis. 
Furthermore, since the basis transformation (\ref{U_beta_action_pos}) is a unitary operation, the effects of geometry-matter entanglement in the first formalism cannot be undone by this change. 
In other words, although the entanglement of states is basis-dependent, and~therefore not fundamental, predictions for the results of physical measurements arise from the combination of both states and operators. 
These predictions are basis-independent, as required \cite{Lake:2019nmn}. 

From Eqs. (\ref{QQ'}) and (\ref{PiPi'}) it is straightforward to show that the new operators $\left\{\hat{Q}^{i},\hat{\Pi}_{i},\hat{Q}'^{i},\hat{\Pi}'_{i}\right\}_{i=1}^3$ satisfy the algebra
\begin{subequations}
\begin{eqnarray} \label{[Q,Pi]}
[\hat{Q}^{i},\hat{\Pi}_{j}] = i\hbar \delta^{i}{}_{j} \, {\bf\hat{\mathbb{I}}} \, , \quad [\hat{Q}'^{i},\hat{\Pi}'_{j}]  = i\beta \delta^{i}{}_{j} \, {\bf\hat{\mathbb{I}}} \, ,
\end{eqnarray}
\begin{eqnarray} \label{mixed_comm-1}
[\hat{Q}^{i},\hat{\Pi}'_{j}] = [\hat{Q}'^{i},\hat{\Pi}_{j}]  = 0 \, ,
\end{eqnarray}
\begin{eqnarray} \label{QQ_commutators}
[\hat{Q}^{i},\hat{Q}^{j}] = [\hat{Q}'^{i},\hat{Q}'^{j}] = 0 \, ,
\end{eqnarray}
\begin{eqnarray} \label{PiPi_commutators}
\quad [\hat{\Pi}_{i}, \hat{\Pi}_{j}] = [\hat{\Pi}'_{i}, \hat{\Pi}'_{j}] = 0 \, ,
\end{eqnarray}
\begin{eqnarray} \label{remaining_commutators}
[\hat{Q}^{i},\hat{Q}'^{j}] = 0 \, , \quad [\hat{\Pi}_{i}, \hat{\Pi}'_{j}] = 0 \, . 
\end{eqnarray}
\end{subequations}
Together, Eqs. (\ref{[Q,Pi]}) and (\ref{mixed_comm-1}) recover Eq. (\ref{[X,P]}) and the remaining commutation relations (\ref{QQ_commutators})-(\ref{remaining_commutators}) recover the rest of the rescaled Heisenberg algebra, Eqs. (\ref{XX_PP_commutators}). 

We then have
\begin{eqnarray} \label{comp-1}
(\Delta_{\Psi}X^{i})^2 &=& (\Delta_{\Psi}Q^{i})^2 + (\Delta_{\Psi}Q'^{i})^2 \, ,
\end{eqnarray}
\begin{eqnarray} \label{comp-2}
(\Delta_{\Psi}P_{j})^2 &=& (\Delta_{\Psi}\Pi_{j})^2 + (\Delta_{\Psi}\Pi'_{j})^2 \, , 
\end{eqnarray}
since ${\rm cov}_{\Psi}(\hat{Q}^{i},\hat{Q}'^{i}) = {\rm cov}_{\Psi}(\hat{Q}'^{i},\hat{Q}^{i}) = 0$ and 
${\rm cov}_{\Psi}(\hat{\Pi}_{j},\hat{\Pi}'_{j}) = {\rm cov}_{\Psi}(\hat{\Pi}'_{j},\hat{\Pi}_{j}) = 0$, where ${\rm cov}(X,Y) = \braket{XY} - \braket{X}\braket{Y}$ denotes the covariance of the random variables $X$ and $Y$. 
The operator pairs $\left\{\hat{Q}^{i},\hat{Q}'^{i}\right\}_{i=1}^{3}$ and $\left\{\hat{\Pi}_{i},\hat{\Pi}'_{i}\right\}_{i=1}^{3}$ are uncorrelated, because they act on separate subspaces of the total state $\ket{\Psi}$, so that the two copies of the shift-isometry algebra commute, as claimed \cite{Lake:2019nmn}. 

Comparing of Eqs. (\ref{comp-1}) and (\ref{comp-2}) with Eqs. (\ref{X_uncertainty}) and (\ref{P_uncertainty}), respectively, shows that
\begin{eqnarray} \label{}
\Delta_{\Psi}Q^{i} = \Delta_{\psi}x'^{i} \, , \quad \Delta_{\Psi}Q'^{i} = \Delta_{g}x'^{i} \, , 
\end{eqnarray}
\begin{eqnarray} \label{}
\Delta_{\Psi}\Pi_{j} = \Delta_{\psi}p'_{j} \, , \quad \Delta_{\Psi}\Pi'_{j} = \Delta_{g}p'_{j} \, .
\end{eqnarray}
We now see the origin of the smeared space GUP (\ref{X_uncertainty}) and EUP (\ref{P_uncertainty}), plus the rescaled commutation relation (\ref{[X,P]}), more clearly. 
One copy of the shift-isometry algebra, scaled by $\hbar$, generates the linear momentum of the matter sector. 
This is equivalent to the canonical position-momentum commutator for a quantum point-particle. 
In addition, the linear momentum carried by `points' in the quantum background geometry is generated by a second copy of this algebra, scaled by $\beta$. 
Since the two representations commute, and, since each copy of the position-momentum commutator is proportional to the same constant, $\delta^{i}{}_{j}$, we are left with a single factor of $(\hbar + \beta)/2 \, . \, \delta^{i}{}_{j}$ on the right-hand side of Eq. (\ref{[X,P]}).

The analysis above shows that the ultimate origin of both the rescaling $\hbar \rightarrow \hbar + \beta$ and the GURs (\ref{comp-1})-(\ref{comp-2}), which are equivalent to the GUP (\ref{X_uncertainty}) and EUP (\ref{P_uncertainty}), respectively, is the generalised algebra (\ref{[Q,Pi]})--(\ref{remaining_commutators}). 
This gives a clearer picture of how GURs may emerge from theories in which Euclidean symmetries are preserved. 
In short, while standard approaches seek to generate GURs by breaking Euclidean invariance \cite{Kempf:1994su,Tawfik:2014zca,Tawfik:2015rva}, we generate them by extending it. 
We apply the Euclidean symmetry algebra to two states in a separable tensor product, $\ket{\Psi} = \ket{\psi} \otimes \ket{g}$ (\ref{Psi_unitary_equiv}). 
The first state represents the canonical quantum matter living on, or `in', the background geometry, while the second state represents the geometry itself. 
The latter is endowed with new, genuinely quantum, degrees of freedom. 
The quantum mechanical amplitudes associated with both subspaces contribute to the total position and momentum uncertainties, as in Eqs. (\ref{comp-1})-(\ref{comp-2}). 

Next, we use the alternative formalism presented here to derive the generalised algebra for smeared angular momentum operators. 
By analogy with our previous results, we show that this corresponds to a simple rescaling of the canonical ${\rm so}(3)$ Lie algebra, such that $\hbar \rightarrow \hbar + \beta$, plus a complex subalgebra structure that gives rise to GURs for angular momentum.

\subsection{Generalised angular momentum measurements} \label{sec:4.2}

As each geometry in the smeared superposition of geometries in intrinsically flat (see Sec. \ref{sec:3.1}), we may construct the generalised displacement and momentum vector operators by analogy with their canonical counterparts as $\hat{{\bf X}} = \hat{X}^i\bold{e}_i(0)$ and $\hat{{\bf P}} = \hat{P}_j\bold{e}^j(0)$, respectively, where $\eta_{ij}(X) = \braket{\bold{e}_i(X),\bold{e}_j(X)} = {\rm diag}(1,1,1)$ and the coordinates $X$ denote global Cartesians. 
These expressions are formally the same as those given in Eq. (\ref{KMM_vector_ops}), for the KMM model, but with $\hat{X}^i$ and $\hat{P}_j$ given by Eqs. (\ref{X_unitary_equiv}) and (\ref{P_unitary_equiv}). 
However, in this case, we can be sure that these definitions are compatible with the generalised commutator (\ref{[X,P]}), since this respects the symmetries of Euclidean space.
\footnote{A priori, we cannot be sure that Eq. (\ref{KMM_vector_ops}) is compatible with the generalised commutator in the KMM case, or in any other modified commutator model. Although the standard assumptions that $\hat{{\bf X}} = \hat{X}^i\bold{e}_i(0)$ and $\hat{{\bf P}} = \hat{P}_j\bold{e}^j(0)$, even when $[\hat{X}^i,\hat{P}_j] \neq {\rm const.} \times \delta^{i}{}_{j}$ \cite{Kempf:1994su,Tawfik:2014zca,Tawfik:2015rva}, may be compatible with each other, this is not clear unless the isometries of the background, and the coordinate axes to which the values $X^i$ and $P_j$ refer, are specified explicitly. Specifying the latter is equivalent to specifying the tangent space structure at each point $X$, which, in turn, is equivalent to specifying the metric geometry of real space \cite{Frankel:1997ec,Nakahara:2003nw}. Potentially, these considerations indicate yet another inconsistency of modified commutator models, and are of particular relevance for generalised models of angular momentum \cite{Bosso:2016frs,Kempf:1994su}. (See \cite{Lake:2019nmn} for further discussion.) For the necessity of quantising in global Cartesians, which only exist in flat space \cite{Frankel:1997ec,Nakahara:2003nw}, see \cite{Messiah:book}.}

The smeared space angular momentum operator is then defined as
\begin{eqnarray} \label{L}
\hat{{\bf L}} = \hat{{\bf X}} \times \hat{{\bf P}} \, . 
\end{eqnarray}
and its Cartesian components are given by
\begin{eqnarray} \label{L_i}
\hat{L}_{i} = (\hat{{\bf X}} \times \hat{{\bf P}})_{i} = \epsilon_{ij}{}^{k}\hat{X}^{j}\hat{P}_{k} \, ,
\end{eqnarray}
where $ \epsilon_{ij}{}^{k}$ is the Levi-Civita symbol. 
This is defined to be $+1$ for symmetric permutations of the indices, $-1$ for antisymmetric permutations, and zero if any two indices are equal \cite{Frankel:1997ec,Nakahara:2003nw}. 
We stress that the operator $\hat{{\bf L}} = \hat{L}_{i}\bold{e}^i(0)$ is well defined by Eqs. (\ref{L})-(\ref{L_i}) only because the vectors ${\bf X} = X^i\bold{e}_i(0)$, ${\bf P} = P_j\bold{e}^j(0)$, and pseudo vector 
${\bf L} = L_{i}\bold{e}^i(0)$, are well defined in Euclidean geometry \cite{LandauMechanics}. 
Potentially, any modification of the Heisenberg algebra, such that $[\hat{X}^i,\hat{P}_j] \neq {\rm const.} \times \delta^{i}{}_{j}$, may be incompatible with constructions of this form \cite{Lake:2019nmn}. 

From Eqs. (\ref{[X,P]}), (\ref{XX_PP_commutators}) and (\ref{L_i}), it follows that
\begin{eqnarray} \label{commutators_LX_LP}
[\hat{L}_{i},\hat{X}^{k}] = i(\hbar + \beta) \, \epsilon_{ij}{}^{k} \hat{X}^{j} \, , \quad [\hat{L}_{i},\hat{P}_{j}] = i(\hbar + \beta) \, \epsilon_{ij}{}^{k} \hat{P}_{k}
\end{eqnarray}
and
\begin{eqnarray} \label{LL_commutator}
[\hat{L}_{i},\hat{L}_{j}] = i(\hbar +\beta) \epsilon_{ij}{}^{k}\hat{L}_{k} \, ,  
\end{eqnarray}
\begin{eqnarray} \label{L^2L_commutator}
[\hat{L}^2,\hat{L}_{i}] = 0 \, .
\end{eqnarray}
Hence, all the operators of the smeared space model, considered so far, respect the symmetry algebras of canonical QM. 
The generalised algebras are equivalent to the canonical ones, but with the simple rescaling $\hbar \rightarrow \hbar + \beta$. 
In particular, Eq. (\ref{LL_commutator}) is simply a representation of the ${\rm so}(3)$ algebra of Euclidean space, which represents rotational invariance \cite{Jones98}. 

Because of this, it is not immediately clear how (or why) smeared space is capable of generating GURs for angular momentum, or for any other observables. 
The key point is that, although the model implies only a simple rescaling of the canonical Schr{\" o}dinger--Robertson bound, for any pair of operators, it nonetheless generates GURs of the form $\Delta_{\Psi}X^{i}\Delta_{\Psi}P_{j} \geq \dots \geq (\hbar + \beta)/2 \, . \, \delta^{i}{}_{j}$, $\Delta_{\Psi}L_{i}\Delta_{\Psi}L_{j} \geq \dots \geq (\hbar + \beta)/2 \, . \, |\epsilon_{ij}{^{k}}\braket{\hat{L}_{k}}_{\Psi}|$, etc. 
The dots in the middle of each of these expressions represent a sum of terms which is generically larger than the Schr{\" o}dinger--Robertson bound on the far right-hand side. 
The resulting hierarchy of inequalities is saturated only under specific conditions, in which the terms in the middle are optimised with respect to the relevant variables.

Therefore, in order to gain deeper insight into the behaviour of angular momentum in the smeared space model, we must investigate the origins of the relations (\ref{commutators_LX_LP})--(\ref{L^2L_commutator}) in more detail. 
We now show explicitly that, despite their almost canonical form (except for the rescaling $\hbar \rightarrow \hbar + \beta$), these expressions are compatible with GURs of the type discussed above. 
In this sense, they are analogous to Eq. (\ref{DXDP_opt}), which, despite its canonical form (except for $\hbar \rightarrow \hbar + \beta$), is compatible with the EGUP (\ref{smeared-spaceEGUP-2}). 

In terms of our new position and momentum operators, (\ref{ops_split_Q}) and (\ref{ops_split_Pi}), the components of the generalised angular momentum may be written as
\begin{eqnarray} \label{L_sum}
\hat{L}_{i} = \hat{\mathcal{L}}_{i} + \hat{\mathcal{L}}'_{i} + \hat{\Lambda}_{i} + \hat{\Lambda}'_{i} \, , 
\end{eqnarray}
where
\begin{eqnarray} \label{L_terms}
\hat{\mathcal{L}}_{i} &=& \epsilon_{ij}{}^{k} \hat{Q}^{j}\hat{\Pi}_{k} \, , \quad \hat{\mathcal{L}}'_{i} = \epsilon_{ij}{}^{k} \hat{Q}'^{j}\hat{\Pi}'_{k} \, ,
\nonumber\\
\hat{\Lambda}_{i} &=& \epsilon_{ij}{}^{k} \hat{Q}^{j}\hat{\Pi}'_{k} \, , \quad \hat{\Lambda}'_{i} = \epsilon_{ij}{}^{k} \hat{Q}'^{j}\hat{\Pi}_{k} \, .
\end{eqnarray}
It is then straightforward to verify that the individual subcomponents of the generalised generators (\ref{L_sum}), $\left\{\hat{\mathcal{L}}_{i},\hat{\mathcal{L}}'_{i},\hat{\Lambda}_{i},\hat{\Lambda}'_{i}\right\}_{i=1}^{3}$, satisfy the algebra

\begin{subequations} \label{rearrange-L}
\begin{equation} \label{rearrange-L.1}
[\hat{\mathcal{L}}_{i},\hat{\mathcal{L}}_{j}] = i\hbar \, \epsilon_{ij}{}^{k} \hat{\mathcal{L}}_{k} \, , \quad [\hat{\mathcal{L}}'_{i},\hat{\mathcal{L}}'_{j}] = i\beta \, \epsilon_{ij}{}^{k} \hat{\mathcal{L}}'_{k} \, , 
\end{equation}
\begin{equation} \label{rearrange-L.2}
[\hat{\mathcal{L}}_{i},\hat{\mathcal{L}}'_{j}] = [\hat{\mathcal{L}}'_{i},\hat{\mathcal{L}}_{j}] = 0 \, , 
\end{equation}
\begin{equation} \label{rearrange-L.3}
[\hat{\mathcal{L}}_{i},\hat{\Lambda}_{j}] - [\hat{\mathcal{L}}_{j},\hat{\Lambda}_{i}] = i\hbar \, \epsilon_{ij}{}^{k} \hat{\Lambda}_{k} \, , 
\end{equation}
\begin{equation} \label{rearrange-L.4}
[\hat{\mathcal{L}}_{i},\hat{\Lambda}'_{j}] - [\hat{\mathcal{L}}_{j},\hat{\Lambda}'_{i}] = i\hbar \, \epsilon_{ij}{}^{k} \hat{\Lambda}'_{k} \, , 
\end{equation}
\begin{equation} \label{rearrange-L.5}
[\hat{\mathcal{L}}'_{i},\hat{\Lambda}_{j}] - [\hat{\mathcal{L}}'_{j},\hat{\Lambda}_{i}] = i\beta \, \epsilon_{ij}{}^{k} \hat{\Lambda}_{k} \, , 
\end{equation}
\begin{equation} \label{rearrange-L.6}
[\hat{\mathcal{L}}'_{i},\hat{\Lambda}'_{j}] - [\hat{\mathcal{L}}'_{j},\hat{\Lambda}'_{i}] = i\beta \, \epsilon_{ij}{}^{k} \hat{\Lambda}'_{k} \, , 
\end{equation}
\begin{equation} \label{rearrange-L.7}
[\hat{\Lambda}_{i},\hat{\Lambda}_{j}] = [\hat{\Lambda}'_{i},\hat{\Lambda}'_{j}] = 0 \, , 
\end{equation}
\begin{equation} \label{rearrange-L.8}
[\hat{\Lambda}_{i},\hat{\Lambda}'_{j}] - [\hat{\Lambda}_{j},\hat{\Lambda}'_{i}] = i\beta \, \epsilon_{ij}{}^{k}\hat{\mathcal{L}}_{k} + i\hbar \, \epsilon_{ij}{}^{m}\hat{\mathcal{L}}'_{m} \, .
\end{equation}
\end{subequations}
Note that summing the left-hand sides of Eqs. (\ref{rearrange-L.1})--(\ref{rearrange-L.8}) yields the generalised commutator $[\hat{L}_{i},\hat{L}_{j}]$ whereas summing the right-hand sides yields $i(\hbar + \beta)\, \epsilon_{ij}{}^{k} \hat{L}_{k}$, as required.

Equations (\ref{rearrange-L.1}) confirm that $\hat{\mathcal{L}}_{i}$ and $\hat{\mathcal{L}}'_{i}$ represent genuine angular momentum operators since the subsets $\left\{\hat{\mathcal{L}}_{i}\right\}_{i=1}^{3}$ and $\left\{\hat{\mathcal{L}}'_{i}\right\}_{i=1}^{3}$ satisfy the required algebras, that is, appropriately scaled representations of ${\rm so}(3)$. 
According to our previous interpretation of the tensor product state (\ref{Psi_unitary_equiv}), $\hat{\mathcal{L}}_{i}$ represents the angular momentum of the canonical quantum state vector $\ket{\psi}$ (quantised at the scale $\hbar$), whereas $\hat{\mathcal{L}}'_{i}$ represents the angular momentum associated with the quantum state of the background $\ket{g}$ (quantised at the scale $\beta$).
By contrast, Eqs. (\ref{rearrange-L.7}) and (\ref{rearrange-L.8}) show that $\hat{\Lambda}_{i}$ and $\hat{\Lambda}'_{i}$ do not represent components of angular momentum in their own right. 
These `cross terms' determine the effect, on the angular momentum of a canonical quantum particle, of its interaction with the smeared background space. 

We also note that, since neither $\hat{\Lambda}_{i}$ nor $\hat{\Lambda}'_{i}$ commute with either $\hat{\mathcal{L}}_{i}$ or $\hat{\mathcal{L}}'_{i}$, it is impossible for a smeared state $\ket{\Psi}$ to be an eigenvector of all four subcomponents of $\hat{L}_{i}$ simultaneously. 
Nonetheless, Eq. (\ref{L^2L_commutator}) demonstrates that the simultaneous eigenvectors of $\hat{L}^2$ and $\hat{L}_{i}$ form a valid basis of the Hilbert space with countably infinite dimensions, $\mathcal{H}$.
Hence, if both $\ket{\psi}$ and $\ket{g}$ are angular momentum eigenstates of their respective operators, i.e., if $\hat{\mathcal{L}}_{i}\ket{\Psi} = m\hbar\ket{\Psi}$ and $\hat{\mathcal{L}}'_{i}\ket{\Psi} = m'\beta\ket{\Psi}$ for some $m,m' \in \mathbb{Z}$, the total state $\ket{\Psi} = \ket{\psi} \otimes \ket{g}$ is not an eigenstate of $\hat{L}_{i}$. 
In this way, single-particle smeared states differ starkly from unentangled bipartite states in canonical QM, $\ket{\psi_{\rm tot}} = \ket{\psi_{1}} \otimes \ket{\psi_{2}}$ \cite{Rae}.
 
Alternatively, we may write the generalised operator $\hat{L}_i$ as
\begin{eqnarray} \label{L_sum*}
\hat{L}_{i} = \hat{\mathcal{L}}_{i} + \hat{\mathcal{L}}'_{i} + \hat{\mathbb{L}}_{i} \, , 
\end{eqnarray}
where
\begin{eqnarray} \label{mathb{L}_{i}}
\hat{\mathbb{L}}_{i} = \hat{\Lambda}_{i} + \hat{\Lambda}'_{i} \, .
\end{eqnarray}
Arguably, this is a more physically relevant decomposition than Eq. (\ref{L_sum}) since it is the sum of terms (\ref{mathb{L}_{i}}) that represents the total interaction of the particle with the background. 
The new subcomponents $\left\{\hat{\mathcal{L}}_{i},\hat{\mathcal{L}}'_{i},\hat{\mathbb{L}}_{i}\right\}_{i=1}^{3}$ then satisfy the algebra
\begin{subequations} \label{rearrange-L*}
\begin{equation} \label{rearrange-L.1*}
[\hat{\mathcal{L}}_{i},\hat{\mathcal{L}}_{j}] = i\hbar \, \epsilon_{ij}{}^{k} \hat{\mathcal{L}}_{k} \, , \quad [\hat{\mathcal{L}}'_{i},\hat{\mathcal{L}}'_{j}] = i\beta \, \epsilon_{ij}{}^{k} \hat{\mathcal{L}}'_{k} \, , 
\end{equation}
\begin{equation} \label{rearrange-L.2*}
[\hat{\mathcal{L}}_{i},\hat{\mathcal{L}}'_{j}] = [\hat{\mathcal{L}}'_{i},\hat{\mathcal{L}}_{j}] = 0 \, , 
\end{equation}
\begin{equation} \label{rearrange-L.3*}
[\hat{\mathcal{L}}_{i},\hat{\mathbb{L}}_{j}] - [\hat{\mathcal{L}}_{j},\hat{\mathbb{L}}_{i}] = i\hbar \, \epsilon_{ij}{}^{k} \hat{\mathbb{L}}_{k} \, , 
\end{equation}
\begin{equation} \label{rearrange-L.4*}
[\hat{\mathcal{L}}'_{i},\hat{\mathbb{L}}_{j}] - [\hat{\mathcal{L}}'_{j},\hat{\mathbb{L}}_{i}] = i\beta \, \epsilon_{ij}{}^{k} \hat{\mathbb{L}}_{k} \, , 
\end{equation}
\begin{equation} \label{rearrange-L.5*}
[\hat{\mathbb{L}}_{i},\hat{\mathbb{L}}_{j}] = i\beta \, \epsilon_{ij}{}^{k}\hat{\mathcal{L}}_{k} + i\hbar \, \epsilon_{ij}{}^{m}\hat{\mathcal{L}}'_{m} \, ,
\end{equation}
\end{subequations}

We note that Eqs. (\ref{rearrange-L.1*})-(\ref{rearrange-L.5*}) are less restrictive than Eqs. (\ref{rearrange-L.1})-(\ref{rearrange-L.8}), in the sense that the former imply the latter, but the latter do not necessitate the former. 
Thus, we may in principle construct an alternative set of operators $\left\{\hat{\mathcal{L}}_{i},\hat{\mathcal{L}}'_{i},\hat{\mathbb{L}}_{i}\right\}_{i=1}^{3}$, which are not defined by Eqs. (\ref{L_terms}) and (\ref{mathb{L}_{i}}), but which nonetheless satisfy the algebra (\ref{rearrange-L*}). 
In this work, we will not investigate alternative solutions of either (\ref{rearrange-L}) or (\ref{rearrange-L*}) in detail. 
However, we note that, keeping our previous definitions of $\hat{\mathcal{L}}_{i}$ and $\hat{\mathcal{L}}'_{i}$, given in Eq. (\ref{L_terms}), and defining the new operators $\hat{\mathbb{L}}_{i} = \frac{2}{\sqrt{\hbar\beta}}\epsilon_{i}{}^{jk}\hat{\mathcal{L}}_{j}\hat{\mathcal{L}}'_{k}$ (*), we may satisfy Eqs. (\ref{rearrange-L.1*})--(\ref{rearrange-L.4*}) but not Eq. (\ref{rearrange-L.5*}).  

The operators (*) are not equivalent to those defined in Eq. (\ref{mathb{L}_{i}}) and do not fully satisfy the algebra (\ref{rearrange-L*}). 
Nonetheless, they offer an important clue about generalised spin physics in the smeared space model, which will be considered in detail in Sec. \ref{sec:4.3}. 
Therein, we show that it is straightforward to construct finite-dimensional analogues of the subcomponents $\left\{\hat{\mathcal{L}}_{i}\right\}_{i=1}^{3}$ and $\left\{\hat{\mathcal{L}}'_{i}\right\}_{i=1}^{3}$. 
However, it it is far less obvious how to construct spin-operator counterparts of the commuting components $\left\{\hat{\Lambda}_i\right\}_{i=1}^{3}$ and $\left\{\hat{\Lambda}'_i\right\}_{i=1}^{3}$. 
Despite this, simple spin-operator analogues of the total interaction terms $\left\{\hat{\mathbb{L}}_{i}\right\}_{i=1}^{3}$ exist. 
These take a form analogous to (*) but with $\hat{\mathcal{L}}_{i}$ and $\hat{\mathcal{L}}'_{i}$ replaced by their finite-dimensional counterparts. 
We then show that, if the spin part of background state $\ket{g}$ is assumed to be fermionic, with eigenvalues $\pm \beta/2$, the resulting generalised spin operators satisfy all the equations of a generalised spin algebra. 
This algebra has the same formal structure as Eqs. (\ref{rearrange-L.1*})-(\ref{rearrange-L.5*}). 
Together, these generate a rescaled ${\rm su}(2)$ algebra for the total spin operators, which act on the composite matter-plus-geometry state, with $\hbar \rightarrow \hbar + \beta$. 
The subalgebra structure also gives rise to GURs for the generalised spin measurements. 

Yet again, we assume that only material degrees of freedom are physically accessible, i.e., that measurements are performed on material bodies in space, but that the geometry is not probed directly.
Nonetheless, the consistency of our model requires spinning particles to interact with the spin of the smeared spatial background, in a way that affects their measured spin values. 
This is formally analogous to the interaction between angular momenta, represented by the algebra (\ref{rearrange-L.1*})-(\ref{rearrange-L.5*}).

However, before considering the case of spin, we demonstrate that the generalised algebras (\ref{rearrange-L.1})-(\ref{rearrange-L.8}) and (\ref{rearrange-L.1*})-(\ref{rearrange-L.5*}) generate GURs for angular momentum. 
Depending on which algebra we choose, the uncertainties of the generalised angular momentum operators $\left\{\hat{L}_{i}\right\}_{i=1}^{3}$ (\ref{L_sum}) may be expressed in terms of the subcomponents $\left\{\hat{\mathcal{L}}_{i},\hat{\mathcal{L}}'_{i},\hat{\Lambda}_{i},\hat{\Lambda}'_{i}\right\}_{i=1}^{3}$ or $\left\{\hat{\mathcal{L}}_{i},\hat{\mathcal{L}}'_{i},\hat{\mathbb{L}}_{i}\right\}_{i=1}^{3}$, respectively. 
In terms of the first set of subcomponents, the variance of an individual component of the generalised angular momentum, $(\Delta_{\Psi}L_{i})^2$, is
\begin{eqnarray} \label{DL^2}
(\Delta_{\Psi}L_{i})^2 &=& (\Delta_{\Psi}\mathcal{L}_{i})^2 + (\Delta_{\Psi}\mathcal{L}'_{i})^2 + (\Delta_{\Psi}\Lambda_{i})^2 + (\Delta_{\Psi}\Lambda'_{i})^2
\nonumber\\
&+& {\rm cov}(\hat{\mathcal{L}}_{i},\hat{\Lambda}_{i}) + {\rm cov}(\hat{\Lambda}_{i},\hat{\mathcal{L}}_{i})
\nonumber\\
&+& {\rm cov}(\hat{\mathcal{L}}_{i},\hat{\Lambda}'_{i}) + {\rm cov}(\hat{\Lambda}'_{i},\hat{\mathcal{L}}_{i})
\nonumber\\
&+& {\rm cov}(\hat{\mathcal{L}}'_{i},\hat{\Lambda}_{i}) + {\rm cov}(\hat{\Lambda}_{i},\hat{\mathcal{L}}'_{i})
\nonumber\\
&+& {\rm cov}(\hat{\mathcal{L}}'_{i},\hat{\Lambda}'_{i}) + {\rm cov}(\hat{\Lambda}'_{i},\hat{\mathcal{L}}'_{i})
\nonumber\\
&+& {\rm cov}(\hat{\Lambda}_{i},\hat{\Lambda}'_{i}) + {\rm cov}(\hat{\Lambda}'_{i},\hat{\Lambda}_{i}) \, ,
\end{eqnarray}
since ${\rm cov}(\hat{\mathcal{L}}_{i},\hat{\mathcal{L}}'_{i}) = {\rm cov}(\hat{\mathcal{L}}'_{i},\hat{\mathcal{L}}_{i}) = 0$. 
The first term on the right-hand side represents the contribution to the total uncertainty from the canonical QM wave function $\psi$, the second represents the pure geometric part (that is, the contribution from $g$), and the additional contributions are generated by operators that cannot be decomposed as either $\hat{\mathbb{I}} \otimes ( \dots )$ or $( \dots ) \otimes \hat{\mathbb{I}}$. 
Equation (\ref{DL^2}) is analogous in form to Eqs. (\ref{comp-1}) and (\ref{comp-2}) but with additional cross terms, i.e., terms generated by operators that do not act on one subspace of the composite state $\ket{\Psi} = \ket{\psi} \otimes \ket{g}$ (\ref{Psi_unitary_equiv}), alone. 

We recall that Eqs. (\ref{comp-1}) and (\ref{comp-2}) are equivalent to Eqs. (\ref{X_uncertainty}) and (\ref{P_uncertainty}) and that these generate the GUP and the EUP, respectively, in the smeared space model. 
In the case of GURs for position and linear momentum, we were able to use a simple theorem about the structure of convolutions to obtain Eqs. (\ref{X_uncertainty}) and (\ref{P_uncertainty}), even when momentum space representation of $\ket{\Psi}$ was expressed in terms of the entangled basis $\ket{{\bf p} \, {\bf p}'}$ (\ref{mod_dB}). 
However, in the case of angular momentum, it was necessary to first express $\ket{\Psi}$ in terms of the separable basis (\ref{Psi_unitary_equiv}) and to define the corresponding `split' operators (\ref{ops_split_Q}) and (\ref{ops_split_Pi}), before the generalised uncertainties $(\Delta_{\Psi}L_{i})^2$ could be decomposed into canonical and non-canonical parts. 
We stress that only the first term on the right-hand side of Eq. (\ref{DL^2}) is present in canonical QM. 
All additional terms are non-canonical and arise as a direct consequence of the smearing procedure (\ref{smearing_map}).

Multiplying Eq. (\ref{DL^2}) by a similar expression for $(\Delta_{\Psi}L_{j})^2$, we obtain the GUR for orbital angular momentum implied by the smeared space model. 
Though it is beyond the scope of this work to investigate the consequences of this relation in detail, we note that it is of the general form 
\begin{eqnarray} \label{L_GUR}
(\Delta_{\Psi}L_{i})^2(\Delta_{\Psi}L_{j})^2 \geq \dots \geq \left(\frac{\hbar + \beta}{2}\right)^2|(\epsilon_{ij}{}^{k})^2\braket{\hat{L}_{k}}_{\Psi}^2| \, ,
\end{eqnarray}
as expected. 
The leading contribution to the terms in the middle is of the form $(\Delta_{\Psi}\mathcal{L}_{i})^2(\Delta_{\Psi}\mathcal{L}_{j})^2 \geq (\hbar/2)^2|(\epsilon_{ij}{}^{k})^2\braket{\hat{\mathcal{L}}_{k}}_{\Psi}^2|$, which is equivalent angular momentum uncertainty relation of canonical QM. 
All other terms represent contributions due to the interaction of the particle with the smeared background. 
In terms of the second set of subcomponents, $(\Delta_{\Psi}L_{i})^2$ may also be written as
\begin{eqnarray} \label{DL^2*}
(\Delta_{\Psi}L_{i})^2 &=& (\Delta_{\Psi}\mathcal{L}_{i})^2 + (\Delta_{\Psi}\mathcal{L}'_{i})^2 + (\Delta_{\Psi}\mathbb{L}_{i})^2 
\nonumber\\
&+& {\rm cov}(\hat{\mathcal{L}}_{i},\hat{\mathbb{L}}_{i}) + {\rm cov}(\hat{\mathbb{L}}_{i},\hat{\mathcal{L}}_{i})
\nonumber\\
&+& {\rm cov}(\hat{\mathcal{L}}'_{i},\hat{\mathbb{L}}_{i}) + {\rm cov}(\hat{\mathbb{L}}_{i},\hat{\mathcal{L}}'_{i}) \, .
\end{eqnarray}
Multiplying by the equivalent expression for $(\Delta_{\Psi}L_{j})^2$, we obtain an alternative (and simpler) form of the GUR for smeared space angular momentum. 
This is still of the general type given by Eq. (\ref{L_GUR}). 

\subsection{Generalised spin measurements} \label{sec:4.3}

To construct a mathematical model of spin measurements in smeared space, we proceed by analogy with the historical development of canonical QM (see \cite{Lake:2019nmn} for details). 
Hence, we seek a set of constant valued matrices $\left\{\hat{S}_{i}\right\}_{i=1}^{3}$ that satisfy the same algebraic structures as the components of angular momentum $\left\{\hat{L}_{i}\right\}_{i=1}^{3}$. 

In the canonical theory, the relevant algebra for the angular momentum operators is simply the three-dimensional rotation algebra, ${\rm so}(3)$, scaled by a factor of $\hbar$. 
However, in the smeared space model, the situation is more complicated. 
In Sec. \ref{sec:4.2}, we showed how the smeared space angular momentum operators can be decomposed into the sum of four terms: a canonical quantum term $\hat{\mathcal{L}}_{i}$ acting on the first subspace of the tensor product state $\ket{\Psi}$ (\ref{Psi_unitary_equiv}), a `pure' geometric part $\hat{\mathcal{L}}'_{i}$ acting on the second subspace, and two cross terms, $\hat{\Lambda}_{i}$ and $\hat{\Lambda}'_{i}$, that act nontrivially on both subspaces (\ref{L_sum})--(\ref{L_terms}). 
The subcomponents $\left\{\hat{\mathcal{L}}_{i},\hat{\mathcal{L}}'_{i},\hat{\Lambda}_{i},\hat{\Lambda}'_{i}\right\}_{i=1}^{3}$ were found to obey the subalgebra defined by Eqs. (\ref{rearrange-L.1})--(\ref{rearrange-L.8}). 
Together, these equations ensure that the rescaled ${\rm so}(3)$ Lie algebra, with $\hbar \rightarrow \hbar + \beta$ (\ref{LL_commutator}), holds for $\left\{\hat{L}_{i}\right\}_{i=1}^{3}$. 
In~addition, we used the alternative definition $\hat{\mathbb{L}}_i = \hat{\Lambda}_{i} +  \hat{\Lambda}'_{i}$ (\ref{mathb{L}_{i}}), leading to the subalgebra (\ref{rearrange-L.1*})--(\ref{rearrange-L.5*}) for $\left\{\hat{\mathcal{L}}_{i},\hat{\mathcal{L}}'_{i},\hat{\mathbb{L}}_i\right\}_{i=1}^{3}$. 

Hence, when searching for generalised spin operators, whose eigenvalues are to be interpreted as the possible spins of the composite matter-plus-geometry quantum state, we have three possible options to explore. 
First, we may search for exact analogues of  Eqs. (\ref{L_sum}) and (\ref{L_terms}). 
This requires that $\hat{S}_{i}$ be decomposed into the sum of four terms, $\hat{S}_{i} = \hat{\mathcal{S}}_{i} + \hat{\mathcal{S}}'_{i} + \hat{\Sigma}_{i} + \hat{\Sigma}'_{i}$, where $\hat{\mathcal{S}}_{i} = \epsilon_{ij}{}^{k}\hat{\alpha}^{j}\hat{\beta}_{k}$, $\hat{\mathcal{S}}'_{i} = \epsilon_{ij}{}^{k}\hat{\alpha}'^{j}\hat{\beta}'_{k}$, $\hat{\Sigma}_{i} = \epsilon_{ij}{}^{k}\hat{\alpha}^{j}\hat{\beta}'_{k}$ and $\hat{\Sigma}'_{i} = \epsilon_{ij}{}^{k}\hat{\alpha}'^{j}\hat{\beta}_{k}$. 
In this case, $\hat{\alpha}^{i}$ and $\hat{\beta}_{j}$ are required to be finite-dimensional constant valued matrices, acting on the first spin-subspace of the smeared tensor product state, that also satisfy the $\hbar$-scaled Heisenberg algebra, i.e., $[\hat{\alpha}^{i},\hat{\beta}_{j}] = i\hbar\delta^{i}{}_{j} \, \hat{\mathbb{I}}$, $[\hat{\alpha}^{i},\hat{\alpha}^{j}] = 0$, $[\hat{\beta}_{i},\hat{\beta}_{j}] = 0$. 
Similarly, $\hat{\alpha}'^{i}$ and $\hat{\beta}'_{j}$ must be finite-dimensional constant valued matrices, acting on the second subspace of the tensor product, that satisfy the $\beta$-scaled Heisenberg algebra $[\hat{\alpha}'^{i},\hat{\beta}'_{j}] = i\beta\delta^{i}{}_{j} \, \hat{\mathbb{I}}$, $[\hat{\alpha}'^{i},\hat{\alpha}'^{j}] = 0$, $[\hat{\beta}'_{i},\hat{\beta}'_{j}] = 0$. 
The requirement that each representation of the Heisenberg algebra acts on a different subspace of the product state also ensures that $[\hat{\alpha}^{i},\hat{\alpha}'^{j}] = 0$, $[\hat{\beta}_{i},\hat{\beta}'_{j}] = 0$, $[\hat{\alpha}_{i},\hat{\beta}'_{j}] = 0$ and $[\hat{\alpha}'_{i},\hat{\beta}_{j}] = 0$. 
(Here, $\hat{\mathbb{I}}$ is used to denote the tensor product of two finite-dimensional subspaces, corresponding to the spins of matter and geometry, respectively.)

However, it is straightforward to show that no such matrices exist. 
The matrices that are most similar to those we require are finite-dimensional representations of the Heisenberg group \cite{HeisenbergGroup}. 
This group has one central element ($z$) and two sets of generators, usually denoted $x^{i}$ and $p_{j}$ by analogy with the canonical commutation relations, that satisfy the algebra $[x^{i},p_{j}] = \delta^{i}{}_{j} \, z$, $[x^{i},x^{j}] = 0$, $[p_{i},p_{j}] = 0$ and $[x^{i},z] = [z,x^{i}]$, $[p_{j},z] = [z,p_{j}]$. 
The central element $z$ commutes with all other generators but, importantly, it does not represent the identity element. 
Confusingly, the commutation relations of the finite-dimensional Heisenberg group are typically referred to as the `Heisenberg algebra' in the mathematical literature, but they are not equivalent to the position-momentum commutation relations of canonical QM \cite{HeisenbergGroup}. 
Therefore, this procedure fails, as it is impossible to define exact finite-dimensional analogues of the subcomponents $\left\{\hat{\mathcal{L}}_{i},\hat{\mathcal{L}}'_{i},\hat{\Lambda}_{i},\hat{\Lambda}'_{i}\right\}_{i=1}^{3}$.

Second, we may search for an alternative set of finite-dimensional constant valued matrices, $\left\{\hat{\mathcal{S}}_{i},\hat{\mathcal{S}}'_{i},\hat{\Sigma}_{i},\hat{\Sigma}'_{i}\right\}_{i=1}^{3}$, that satisfy an algebra analogous to Eqs. (\ref{rearrange-L.1})-(\ref{rearrange-L.8}) under the interchange $\hat{\mathcal{S}}_{i} \leftrightarrow \hat{\mathcal{L}}_{i}$, $\hat{\mathcal{S}}'_{i} \leftrightarrow \hat{\mathcal{L}}'_{i}$, $\hat{\Sigma}_{i} \leftrightarrow \hat{\Lambda}_{i}$ and $\hat{\Sigma}'_{i} \leftrightarrow \hat{\Lambda}'_{i}$. 
By the argument above, these cannot be defined in terms of finite-dimensional analogues of the canonical position and momentum operators, i.e., $\hat{\alpha}^{i} \sim \hat{x}^{i}$, $\hat{\alpha}'^{i} \sim \hat{x}'^{i}$, $\hat{\beta}_{j} \sim \hat{p}_{j}$ and $\hat{\beta}'_{j} \sim \hat{p}'_{j}$. 
In this case, we must again require that $\hat{\mathcal{S}}_{i}$ act nontrivially only on the first subspace of the tensor product state, that $\hat{\mathcal{S}}'_{i}$ act nontrivially only on the second subspace, and that $\hat{\Sigma}_{i}$ and $\hat{\Sigma}'_{i}$ act nontrivially on both subspaces. 

These conditions are very difficult to satisfy. 
The most natural operators that are able to act nontrivially on both spin subspaces are of the form $\hat{\Sigma}_{i} \sim \hat{\Sigma}'_{i} \sim \epsilon_{i}{}^{jk}\sigma_{j} \otimes \sigma'_{k}$. 
However, using these definitions, it is straightforward to show that $[\hat{\Sigma}_{i}, \hat{\Sigma}_{j}] \neq 0$ and $[\hat{\Sigma}'_{i}, \hat{\Sigma}'_{j}] \neq 0$, so that the analogues of Eqs. (\ref{rearrange-L.7}) cannot be satisfied. 
Therefore, this procedure also fails. 

Third, we may search for a smaller set of finite-dimensional constant valued matrices $\left\{\hat{\mathcal{S}}_{i},\hat{\mathcal{S}}'_{i},\hat{\mathbb{S}}_{i}\right\}_{i=1}^{3}$ that satisfy an analogue of the algebra (\ref{rearrange-L.1*})-(\ref{rearrange-L.5*}) under the exchange $\hat{\mathcal{S}}_{i} \leftrightarrow \hat{\mathcal{L}}_{i}$, $\hat{\mathcal{S}}'_{i} \leftrightarrow \hat{\mathcal{L}}'_{i}$ and $\hat{\mathbb{S}}_{i} \leftrightarrow \hat{\mathbb{L}}_{i}$. 
As well as being mathematically simpler, this scenario is also the physically most intuitive. 
In this case, the subcomponents $\left\{\hat{\mathcal{S}}_{i}\right\}_{i=1}^{3}$ determine the ${\rm SU}(2)$ symmetry of canonical quantum matter whereas $\left\{\hat{\mathcal{S}}'_{i}\right\}_{i=1}^{3}$ determine the ${\rm SU}(2)$ symmetry of the quantum state associated with the background. 
The subcomponents $\left\{\hat{\mathbb{S}}_{i}\right\}_{i=1}^{3}$ then determine the spin-spin interaction between canonical QM particles and the smeared geometry. 

Considering the arguments above, we define the generalised spin operator for the composite matter-plus-geometry state, $\hat{S}_i$, as
\begin{eqnarray} \label{S_sum}
\hat{S}_{i} = \hat{\mathcal{S}}_{i} + \hat{\mathcal{S}}'_{i} + \hat{\mathbb{S}}_{i} \, , 
\end{eqnarray}
where $\hat{\mathcal{S}}_{i}$ and $\hat{\mathcal{S}}'_{i}$ are given by
\begin{eqnarray} \label{ss'}
\hat{\mathcal{S}}_{i} = \hat{s}_{i} \otimes \hat{\mathbb{I}}' \, , \quad \hat{\mathcal{S}}'_{i} = \hat{\mathbb{I}} \otimes \hat{s}'_{i} \, , 
\end{eqnarray} 
and
\begin{eqnarray} \label{s'}
\hat{s}_{i} = \frac{\hbar}{2} \, \sigma_{i} \, , \quad \hat{s}'_{i} = \frac{\beta}{2} \, \sigma'_{i} \, .
\end{eqnarray}
The prime on the Pauli operators acting on the second spin-subspace, which corresponds to the spin part of the quantum state associated with the background geometry, indicates that this may posses a different fundamental spin 
to the matter component, $s' \neq s$. 
In this case, the two spin subspaces have different dimensions. 
From here on, we use the shorthand notations $\sigma_{i} = \sigma_{i}(s)$, $\hat{\mathbb{I}} = \hat{\mathbb{I}}_{2s+1}$ and $\sigma'_{i} = \sigma_{i}(s')$, $\hat{\mathbb{I}}' = \hat{\mathbb{I}}_{2s'+1}$. 
It follows from the definitions (\ref{S_sum})--(\ref{s'}) that
\begin{equation} \label{rearrange-S.1*}
[\hat{\mathcal{S}}_{i},\hat{\mathcal{S}}_{j}] = i\hbar \, \epsilon_{ij}{}^{k} \hat{\mathcal{S}}_{k} \, ,  \quad [\hat{\mathcal{S}}'_{i},\hat{\mathcal{S}}'_{j}] = i\beta \, \epsilon_{ij}{}^{k} \hat{\mathcal{S}}'_{k} \, , 
\end{equation}
and
\begin{equation} \label{rearrange-S.2*}
[\hat{\mathcal{S}}_{i},\hat{\mathcal{S}}'_{j}] = [\hat{\mathcal{S}}'_{i},\hat{\mathcal{S}}_{j}] = 0 \, ,
\end{equation}
for any $s, \, s' \in m/2$, $m \in \mathbb{N}$.

Next, we define the interaction term $\hat{\mathbb{S}}_{i}$ as
\begin{eqnarray} \label{mathbb{S}_{i}}
\hat{\mathbb{S}}_{i} = \frac{\sqrt{\hbar\beta}}{2} \, \epsilon_{i}{}^{jk}\sigma_{j} \otimes \sigma'_{k} = \frac{2}{\sqrt{\hbar\beta}}  \, \epsilon_{i}{}^{jk} \hat{\mathcal{S}}_{j}\hat{\mathcal{S}}'_{k} \, . 
\end{eqnarray}
This is the analogue of the operator (*) introduced below Eqs. (\ref{rearrange-L.1*})-(\ref{rearrange-L.5*}) in Sec. \ref{sec:4.2}. 
Using the identity $[AB,C] = A[B,C] + [A,C]B$, Eqs. (\ref{rearrange-S.1*})--(\ref{mathbb{S}_{i}}) are sufficient to show that the relations
\begin{equation} \label{rearrange-S.3*}
[\hat{\mathcal{S}}_{i},\hat{\mathbb{S}}_{j}] - [\hat{\mathcal{S}}_{j},\hat{\mathbb{S}}_{i}] = i\hbar \, \epsilon_{ij}{}^{k} \hat{\mathbb{S}}_{k} \, , 
\end{equation}
\begin{equation} \label{rearrange-S.4*}
[\hat{\mathcal{S}}'_{i},\hat{\mathbb{S}}_{j}] - [\hat{\mathcal{S}}'_{j},\hat{\mathbb{S}}_{i}] = i\beta \, \epsilon_{ij}{}^{k} \hat{\mathbb{S}}_{k} \, , 
\end{equation}
also hold for any $s, \, s' \in m/2$, $m \in \mathbb{N}$. 
Hence, in order to recover a rescaled ${\rm su}(2)$ Lie algebra for the generalised operators $\left\{\hat{S}_i\right\}_{i=1}^{3}$ (with $\hbar \rightarrow \hbar + \beta$), we require the following commutation relations to hold between the cross terms 
$\hat{\mathbb{S}}_{i}$ and $\hat{\mathbb{S}}_{j}$: 
\begin{equation} \label{rearrange-S.5*}
[\hat{\mathbb{S}}_{i},\hat{\mathbb{S}}_{j}] = i\beta \, \epsilon_{ij}{}^{k}\hat{\mathcal{S}}_{k} + i\hbar \, \epsilon_{ij}{}^{m}\hat{\mathcal{S}}'_{m} \, .
\end{equation}

In this work, our main aim is to describe the generalised spin physics of fundamental fermions (e.g., electrons) in smeared space. 
Therefore, although the Standard Model admits composite particles with spin $3/2$ \cite{Donoghue:1992dd}, we restrict our attention to the situation of greatest physical interest and set $s = 1/2$ from here on. 
The operators $\left\{\hat{\mathcal{S}}_{i}\right\}_{i=1}^{3}$ then satisfy the Clifford algebra
\begin{eqnarray} \label{ss_anti-commutator*}
[\hat{\mathcal{S}}_{i},\hat{\mathcal{S}}_{j}]_{+} = \frac{\hbar^2}{2} \, \delta_{ij} \, \hat{\mathbb{I}} \, ,
\end{eqnarray}
where $[ \, . \, , \, . \, ]_{+}$ denotes the anti-commutator, in addition to the Lie algebra given in Eq. (\ref{rearrange-S.1*}) \cite{CliffordAlgebras}. 
These can be combined into the fundamental relation \cite{LieAlgebrasGutowski}
\begin{eqnarray} \label{fundamental_relation_spin_ops*}
\hat{\mathcal{S}}_{i}\hat{\mathcal{S}}_{j} = \left(\frac{\hbar}{2}\right)^2 \, \delta_{ij} \hat{\mathbb{I}}  + \, i\left(\frac{\hbar}{2}\right)\epsilon_{ij}{}^{k}\hat{\mathcal{S}}_{k} \, . 
\end{eqnarray}
It is then straightforward to show that Eq. (\ref{rearrange-S.5*}) holds if
\begin{eqnarray} \label{ss_anti-commutator**}
[\hat{\mathcal{S}}'_{i},\hat{\mathcal{S}}'_{j}]_{+} = \frac{\beta^2}{2} \, \delta_{ij} \, \hat{\mathbb{I}} \, ,
\end{eqnarray}
which, together with the $\beta$-scaled Lie algebra in Eq. (\ref{rearrange-S.1*}), implies
\begin{eqnarray} \label{fundamental_relation_spin_ops**}
\hat{\mathcal{S}}'_{i}\hat{\mathcal{S}}'_{j} = \left(\frac{\beta}{2}\right)^2 \, \delta_{ij} \hat{\mathbb{I}}  + \, i\left(\frac{\beta}{2}\right)\epsilon_{ij}{}^{k}\hat{\mathcal{S}}'_{k} \, . 
\end{eqnarray}
We stress that, unlike Eqs. (\ref{rearrange-S.1*})--(\ref{rearrange-S.2*}) and (\ref{rearrange-S.3*})--(\ref{rearrange-S.4*}), the relations (\ref{ss_anti-commutator**}) and (\ref{fundamental_relation_spin_ops**}) hold only when $s' = 1/2$ \cite{LieAlgebrasGutowski,KramersMethod}. 
Consistency of the generalised spin structure therefore implies that the quantum state associated with the background geometry must be fermionic in nature, with spin eigenvalues $\pm\beta/2$.  

In summary, the generalised spin algebra for the whole set of subcomponents $\left\{\hat{\mathcal{S}}_{i},\hat{\mathcal{S}}'_{i},\hat{\mathbb{S}}_{i}\right\}_{i=1}^{3}$ is
\begin{subequations} \label{rearrange-S}
\begin{equation} \label{rearrange-S.1}
[\hat{\mathcal{S}}_{i},\hat{\mathcal{S}}_{j}] = i\hbar \, \epsilon_{ij}{}^{k} \hat{\mathcal{S}}_{k} \, ,  \quad [\hat{\mathcal{S}}'_{i},\hat{\mathcal{S}}'_{j}] = i\beta \, \epsilon_{ij}{}^{k} \hat{\mathcal{S}}'_{k} \, , 
\end{equation}
\begin{equation} \label{rearrange-S.2}
[\hat{\mathcal{S}}_{i},\hat{\mathcal{S}}'_{j}] = [\hat{\mathcal{S}}'_{i},\hat{\mathcal{S}}_{j}] = 0 \, , 
\end{equation}
\begin{equation} \label{rearrange-S.3}
[\hat{\mathcal{S}}_{i},\hat{\mathbb{S}}_{j}] - [\hat{\mathcal{S}}_{j},\hat{\mathbb{S}}_{i}] = i\hbar \, \epsilon_{ij}{}^{k} \hat{\mathbb{S}}_{k} \, , 
\end{equation}
\begin{equation} \label{rearrange-S.4}
[\hat{\mathcal{S}}'_{i},\hat{\mathbb{S}}_{j}] - [\hat{\mathcal{S}}'_{j},\hat{\mathbb{S}}_{i}] = i\beta \, \epsilon_{ij}{}^{k} \hat{\mathbb{S}}_{k} \, , 
\end{equation}
\begin{equation} \label{rearrange-S.5}
[\hat{\mathbb{S}}_{i},\hat{\mathbb{S}}_{j}] = i\beta \, \epsilon_{ij}{}^{k}\hat{\mathcal{S}}_{k} + i\hbar \, \epsilon_{ij}{}^{m}\hat{\mathcal{S}}'_{m} \, ,
\end{equation}
\end{subequations}
This is formally analogous to the generalised angular momentum algebra, Eqs. (\ref{rearrange-L.1})-(\ref{rearrange-L.5}), under the exchange $\hat{\mathcal{L}}_{i} \leftrightarrow \hat{\mathcal{S}}_{i}$, $\hat{\mathcal{L}}'_{i} \leftrightarrow \hat{\mathcal{S}}'_{i}$ and $\hat{\mathbb{L}}_{i} \leftrightarrow \hat{\mathbb{S}}_{i}$. 
Together, Eqs. (\ref{rearrange-S.1})--(\ref{rearrange-S.5}) give rise to the rescaled ${\rm su}(2)$ Lie algebra
\begin{eqnarray} \label{SS_commutator}
[\hat{S}_{i},\hat{S}_{j}] = i(\hbar + \beta) \epsilon_{ij}{}^{k}\hat{S}_{k} \, ,
\end{eqnarray}
and the rescaled Clifford algebra
\begin{eqnarray} \label{SS_anticommutator}
[\hat{S}_{i},\hat{S}_{j}]_{+} = \frac{(\hbar + \beta)^2}{2} \delta_{ij} \, \hat{\mathbb{I}} \, ,
\end{eqnarray}
for the generalised spin-measurement operators $\left\{\hat{S}_{i}\right\}_{i=1}^{3}$ (\ref{S_sum}). 
From Eq. (\ref{SS_commutator}), plus the definition of the spin vector operator $\hat{\bold{S}} = \hat{S}_{i}\bold{e}^{i}(0)$, it also follows that
\begin{eqnarray} \label{}
[\hat{S}^{2},\hat{S}_{i}] = 0 \, . 
\end{eqnarray}

Note that, in the limit $\hbar \rightarrow \beta$, the spin-spin interaction term $\hat{\mathbb{S}}_{i}$ is not necessary to maintain the canonical Lie algebra structure. 
Since both $\left\{\hat{\mathcal{S}}_{i}\right\}_{i=1}^{3}$ and $\left\{\hat{\mathcal{S}}'_{i}\right\}_{i=1}^{3}$ are representations of the ${\rm su}(2)$ generators, and these representations commute with each other (\ref{ss'}), the combination 
$\hat{\mathcal{S}}_{i} + \hat{\mathcal{S}}'_{i}$ also satisfies the ${\rm su}(2)$ algebra {\it if} both sets of generators are weighted by the same scale factor. 
In this case, we may pull a single factor of $\hbar$ outside the sum of terms on right-hand side of the commutation relations, so that $[\hat{\mathcal{S}}_{i} + \hat{\mathcal{S}}'_{i},\hat{\mathcal{S}}_{j} + \hat{\mathcal{S}}'_{j}] = i \hbar\epsilon_{ij}{}^{k}(\hat{\mathcal{S}}_{k} + \hat{\mathcal{S}}'_{k})$. 
However, after introducing a second quantisation scale for the background, $\beta \neq \hbar$, which is an essential feature of the smeared space model \cite{Lake:2018zeg,Lake:2019nmn}, the interaction represented by $\hat{\mathbb{S}}_{i}$ is necessary to maintain the ${\rm su}(2)$ algebra of the composite matter-plus-geometry system. 
Without this interaction it is not possible to construct a generalised operator $\hat{S}_{i}$, that includes commuting representations of the $i^{\rm th}$ ${\rm SU}(2)$ generator weighted by different scale factors, i.e., $\hat{\mathcal{S}}_{i} = (\hbar/2) (\sigma_{i} \otimes \mathbb{I}')$ and $\hat{\mathcal{S}}'_{i} = (\beta/2) (\mathbb{I} \otimes \sigma'_{i})$ with $\beta \neq \hbar$, and which also satisfies a canonical-type commutation relation. 
In this case, it is not possible to pull a single factor (with units of action) outside the expression on the right-hand side of the relation $[\hat{S}_{i},\hat{S}_{j}] = (\dots)$, without including $\hat{\mathbb{S}}_{i}$ (\ref{mathbb{S}_{i}}) in the definition of $\hat{S}_{i}$ (\ref{S_sum}). 

This is a fundamental difference between canonical two-particle states and the bipartite matter-plus-geometry states of the smeared space model \cite{Lake:2019nmn,Rae}. 
Furthermore, it has clear physical interpretation. 
The first copy of the ${\rm su}(2)$ algebra, weighted by $\hbar$, defines the isometry that generates the spin of the matter sector. 
The second copy, weighted by $\beta$, generates the quantum spin of the background. 
The introduction of a second quantisation scale for geometry, $\beta \neq \hbar$, breaks the ${\rm SU(2)}$ invariance of the composite matter-plus-geometry state. 

If the spins of matter and geometry are left to evolve freely, without interacting, this violation of ${\rm SU(2)}$ symmetry is manifested by violation of the corresponding Lie algebra, ${\rm su(2)}$. 
However, the spins do not evolve freely, but interact via the cross term $\hat{\mathbb{S}}_{i}$. 
The interaction is such that the ${\rm su}(2)$ algebra is restored, for the composite state, under the simple rescaling $\hbar \rightarrow \hbar + \beta$. 
Analogous statements also hold for the generalised angular momentum operators $\left\{\hat{L}_{i}\right\}_{i=1}^{3}$, with ${\rm SU(2)}$ and ${\rm su(2)}$ replaced by ${\rm SO(3)}$ and ${\rm so(3)}$, respectively. 
This gives a clear picture of the physical effects of the interaction terms $\left\{\hat{\mathbb{L}}_{i}\right\}_{i=1}^{3}$.

Written explicitly, the generalised spin matrices take the form
\begin{eqnarray} \label{S_i_explicit}
&&\hat{S}_{x} = 
\begin{bmatrix}
    0  &  \frac{(\beta + i\sqrt{\hbar\beta})}{2} & \frac{(\hbar - i\sqrt{\hbar\beta})}{2}  &  0 \\
    \frac{(\beta - i\sqrt{\hbar\beta})}{2}  &  0 & 0  &  \frac{(\hbar + i\sqrt{\hbar\beta})}{2} \\
    \frac{(\hbar + i\sqrt{\hbar\beta})}{2} &  0 & 0  &  \frac{(\beta - i\sqrt{\hbar\beta})}{2} \\
    0  &  \frac{(\hbar - i\sqrt{\hbar\beta})}{2}  & \frac{(\beta + i\sqrt{\hbar\beta})}{2} &  0 
\end{bmatrix}
\, , \nonumber\\ 
&&\hat{S}_{y} = 
\begin{bmatrix}
    0  &  -\frac{(i\beta - \sqrt{\hbar\beta})}{2} & -\frac{(i\hbar + \sqrt{\hbar\beta})}{2}  &  0 \\
    \frac{(i\beta + \sqrt{\hbar\beta})}{2}  &  0 & 0  &  -\frac{(i\hbar - \sqrt{\hbar\beta})}{2} \\
    \frac{(i\hbar - \sqrt{\hbar\beta})}{2} &  0 & 0  &  -\frac{(i\beta + \sqrt{\hbar\beta})}{2} \\
    0  &  \frac{(i\hbar + \sqrt{\hbar\beta})}{2}  & \frac{(i\beta - \sqrt{\hbar\beta})}{2}  &  0 
\end{bmatrix}
\, , \nonumber\\
&&\hat{S}_{z} =
\begin{bmatrix}
    \frac{(\hbar + \beta)}{2}  &  0  &  0  &  0 \\
    0  &  \frac{(\hbar - \beta)}{2}  &  i\sqrt{\hbar\beta}  &  0 \\
    0  &  -i\sqrt{\hbar\beta}  &  -\frac{(\hbar - \beta)}{2}  &  0 \\
    0  &  0  &  0  &  -\frac{(\hbar+\beta)}{2} 
\end{bmatrix}
\, ,
\end{eqnarray}
and $\hat{S}^2$ is given by
\begin{eqnarray}  \label{S^2_explicit}
\hat{S}^{2} = \frac{3(\hbar + \beta)^2}{4} \, \hat{\mathbb{I}}_4 \, , 
\end{eqnarray}
where $\hat{\mathbb{I}}_4$ denotes the four-dimensional identity matrix. 
Equation (\ref{S^2_explicit}) follows from the fact that the matrices $\left\{\left(\frac{\hbar + \beta}{2}\right)^{-1}\hat{S}_{i}\right\}_{i=1}^{3}$ are involutions. 
Hence, in the smeared space model, $\left\{\left(\frac{\hbar + \beta}{2}\right)^{-1}\hat{S}_{i}\right\}_{i=1}^{3}$ are the analogues of the canonical spin-$1/2$ Pauli matrices, $\left\{\sigma_{i}\right\}_{i=1}^{3} = \left\{\left(\frac{\hbar}{2}\right)^{-1}\hat{s}_{i}\right\}_{i=1}^{3}$. 
However, unlike the canonical Pauli matrices, $\left\{\left(\frac{\hbar + \beta}{2}\right)^{-1}\hat{S}_{i}\right\}_{i=1}^{3}$ depend explicitly on both quantisation scales, $\hbar$ and $\beta$.

It is straightforward to show that the simultaneous eigenvectors of $\hat{S}^2$ and $\hat{S}_{i}$, for all three operators in Eqs. (\ref{S_i_explicit}), have eigenvalues
\begin{eqnarray}
\left\{ \frac{(\hbar + \beta)}{2} ,  \frac{(\hbar + \beta)}{2} ,  -\frac{(\hbar + \beta)}{2} ,  -\frac{(\hbar + \beta)}{2} \right\} \, . 
\end{eqnarray}
For $\hat{S}_z$, these correspond to the (un-normalised) eigenvectors
\begin{eqnarray}
\left\{(1,0,0,0),\left(0,\frac{i\hbar}{\sqrt{\hbar\beta}},1,0\right),\left(0,-\frac{i\beta}{\sqrt{\hbar\beta}},1,0\right), (0,0,0,1)\right\} \, , 
\nonumber
\end{eqnarray}
respectively. 
The normalised eigenvectors of $\hat{S}_{z}$ may then be written as
\begin{eqnarray} \label{unentangled_eigenvectors}
\Big|\frac{3(\hbar + \beta)^2}{4}, \left(\frac{\hbar + \beta}{2}\right)_{z}\Big\rangle = (1,0,0,0) = \ket{\uparrow_{z}}_1 \ket{\uparrow_{z}}_2 \, ,  
\nonumber\\ 
\Big|\frac{3(\hbar + \beta)^2}{4}, -\left(\frac{\hbar + \beta}{2}\right)_{z}\Big\rangle = (0,0,0,1) = \ket{\downarrow_{z}}_1 \ket{\downarrow_{z}}_2  \, ,  
\end{eqnarray}
and
\begin{eqnarray} \label{entangled_eigenvectors}
\Big|\frac{3(\hbar + \beta)^2}{4}, \left(\frac{\hbar + \beta}{2}\right)_{z}\Big\rangle_{\delta} &=& \frac{1}{\sqrt{1 + \delta}} (0,1,-i\sqrt{\delta},0)
\nonumber\\
&=& \frac{1}{\sqrt{1 + \delta}}(\ket{\uparrow_{z}}_1 \ket{\downarrow_{z}}_2 - i\sqrt{\delta}\ket{\downarrow_{z}}_1 \ket{\uparrow_{z}}_2) \, , 
\nonumber\\
\Big|\frac{3(\hbar + \beta)^2}{4},  -\left(\frac{\hbar + \beta}{2}\right)_{z}\Big\rangle_{\delta} &=& \frac{1}{\sqrt{1 + \delta}} (0,-i\sqrt{\delta},1,0) 
\nonumber\\ 
&=& \frac{1}{\sqrt{1 + \delta}} (\ket{\downarrow_{z}}_1 \ket{\uparrow_{z}}_2 - i\sqrt{\delta}\ket{\uparrow_{z}}_1 \ket{\downarrow_{z}}_2)  \, ,
\end{eqnarray}
where we have introduced the dimensionless parameter
\begin{eqnarray}
\delta = \hbar/\beta \simeq 10^{-61} \, . 
\end{eqnarray}

The single-electron-plus-smeared-background system has four spin states, as opposed to the two spin states of electrons in the fixed background of canonical QM. 
However, the operators $\hat{S}^{2}$ and $\hat{S}_{i}$ that act on the composite system have only two distinct sets of eigenvalues, $\left\{3(\hbar+\beta)^2/4,\pm(\hbar+\beta)/2\right\}$, and each set has a $2$-fold degeneracy. 
For $\hat{S}_{z}$, each eigenvalue corresponds to one separable state and one state in which the $z$-spins of the electron and the background are entangled.  
The eigenvectors $\ket{\frac{3(\hbar + \beta)^2}{4}, \frac{(\hbar + \beta)}{2}}$ and $\ket{\frac{3(\hbar + \beta)^2}{4}, \frac{(\hbar + \beta)}{2}}_{\delta}$ correspond to spin `up' states, according to their measured eigenvalues, whereas $\ket{\frac{3(\hbar + \beta)^2}{4}, -\frac{(\hbar + \beta)}{2}}$ and $\ket{\frac{3(\hbar + \beta)^2}{4}, -\frac{(\hbar + \beta)}{2}}_{\delta}$ correspond to spin `down' states. 

For the unentangled states, $\ket{\frac{3(\hbar + \beta)^2}{4}, \pm \frac{(\hbar + \beta)}{2}}$, the $z$-spins of the matter and geometry components of the tensor product smeared state, $\ket{\psi}$ and $\ket{g}$, are aligned. 
The spin up state is characterised by the individual values $\left\{+\hbar/2,+\beta/2\right\}$ and the spin down state is characterised by the values $\left\{-\hbar/2,-\beta/2\right\}$. 
However, for the entangled eigenvectors, $\ket{\frac{3(\hbar + \beta)^2}{4}, \pm \frac{(\hbar + \beta)}{2}}_{\delta}$, there is no simple relation between the matter and geometry components of the total quantum state. 
Remarkably, the entangled eigenstates (\ref{entangled_eigenvectors}) have the same eigenvalues as the simple separable states (\ref{unentangled_eigenvectors}). 

We also note that, in the absence the interaction term $\mathbb{S}_{i}$, the eigenvalues of the composite operator $\hat{\mathcal{S}}_{i} + \hat{\mathcal{S}}'_{i}$ are $\left\{(\hbar+\beta)/2,(\hbar-\beta)/2,-(\hbar-\beta)/2,-(\hbar+\beta)/2\right\}$. 
These correspond to the eigenvectors $\left\{\ket{\uparrow_{i}}_1 \ket{\uparrow_{i}}_2,\ket{\uparrow_{i}}_1 \ket{\downarrow_{i}}_2,\ket{\downarrow_{i}}_1 \ket{\uparrow_{i}}_2,\ket{\downarrow_{i}}_1 \ket{\downarrow_{i}}_2\right\}$, respectively, which in the limit $\beta \rightarrow \hbar$ yield the familiar basis vectors of a canonical two-particle state \cite{Rae}.  
Thus, the introduction of $\mathbb{S}_{i}$ not only restores the ${\rm su}(2)$ algebra of the composite matter-plus-geometry system, in the presence of two quantisation scales, $\hbar$ and $\beta \neq \hbar$, but also alters two of the four spin eigenstates of the decoupled sectors, while leaving the remaining two unchanged. 
This, in turn, shifts the corresponding eigenvalues by just the right amount to introduce a $2$-fold degeneracy in the measured values of $\hat{S}^{2}$ and $\hat{S}_{i}$. 

A priori, there was no reason for us to anticipate that the additional terms required to restore the ${\rm su}(2)$ algebra for the generalised operators $\left\{\hat{S}_{i}\right\}_{i=1}^{3}$, i.e., those involving $\mathbb{S}_{i}$ in the subalgebra (\ref{rearrange-S.1})-(\ref{rearrange-S.5}), would simultaneously introduce such a degeneracy.  
However, if had this not occurred, the doubling of the spin degrees of freedom would, in principle, have been directly detectable by performing simultaneous measurements of $\hat{S}^2$ and $\hat{S}_{i}$. 
This would have caused severe problems for the smeared space model, at least philosophically, even if the mathematical formalism remained consistent. 
It is not hard to see why. 
In the non-spin part of the model, the doubling of the canonical degrees of freedom is detectable only indirectly, via the additional statistical fluctuations it induces in the measured values of position, momentum and angular momentum, etc. 
These generate the GURs derived in Secs. \ref{sec:3.1}-\ref{sec:3.3} and \ref{sec:4.2}, which are consistent with our general assumptions about the physical measurement scheme. 
We repeat that we assume a scheme in which measurements are performed only on material bodies in space \cite{Lake:2018zeg,Lake:2019nmn}. 
Therefore, we do not have direct physical access to the quantum degrees of freedom of the background, which can be detected only indirectly via their influence on quantum particles. 

In the first formalism of the model, given in Sec. \ref{sec:3}, this is expressed by tracing out, or, equivalently, integrating out the degrees of freedom in the first subspace of the tensor product Hilbert space, as in Eq. (\ref{EQ_XPRIMEDENSITY}). 
However, in the second formalism, on which the treatment of angular momentum is based, the mathematical structure that renders only half of the doubled phase space directly measurable is more complicated. 
The treatment of spin, given here, is based on this second formalism, but our inability to define finite-dimensional analogues of the canonical QM operators, $\hat{\alpha}^{i} \sim \hat{x}^{i}$, $\hat{\alpha}'^{i} \sim \hat{x}'^{i}$, $\hat{\beta}_{j} \sim \hat{p}_{j}$ and $\hat{\beta}'_{j} \sim \hat{p}'_{j}$, means that there is no clear analogue of either structure in the finite-dimensional case. 
Simply tracing out half of the doubled spin degrees of freedom would require us to make an arbitrary choice, namely, as to which two of the four possible spin states we should regard as physical. 

Remarkably, the algebra (\ref{rearrange-S.1})-(\ref{rearrange-S.5}) saves us from this dilemma, just as it saves the ${\rm su}(2)$ algebra of the composite state in the two-scale quantisation scheme. 
The resulting model of generalised spin measurements is mathematically consistent, and is also consistent with the physical assumptions underlying the smeared space model as a whole, despite the doubling of the number of dimensions in the spin Hilbert space. 
In this case, the doubling is real, since $\mathcal{H}_2 \otimes \mathcal{H}_2 \cong \mathcal{H}_4$, unlike the infinite-dimensional case in which $\mathcal{H} \otimes \mathcal{H} \cong \mathcal{H}$ \cite{HilbertSpaces}. 

Finally, we may write down the GURs implied by the generalised spin algebra (\ref{rearrange-S.1})-(\ref{rearrange-S.5}). 
By analogy with Eq. (\ref{DL^2*}), $(\Delta_{\Psi}S_{i})^2$ takes the form
\begin{eqnarray} \label{DS^2*}
(\Delta_{\Psi}S_{i})^2 &=& (\Delta_{\Psi}\mathcal{S}_{i})^2 + (\Delta_{\Psi}\mathcal{S}'_{i})^2 + (\Delta_{\Psi}\mathbb{S}_{i})^2 
\nonumber\\
&+& {\rm cov}(\hat{\mathcal{S}}_{i},\hat{\mathbb{S}}_{i}) + {\rm cov}(\hat{\mathbb{S}}_{i},\hat{\mathcal{S}}_{i})
\nonumber\\
&+& {\rm cov}(\hat{\mathcal{S}}'_{i},\hat{\mathbb{S}}_{i}) + {\rm cov}(\hat{\mathbb{S}}_{i},\hat{\mathcal{S}}'_{i}) \, .
\end{eqnarray}
Multiplying by the equivalent expression for $(\Delta_{\Psi}S_{j})^2$, we obtain the GUR for spin measurements in smeared space. 
Again, it is beyond the scope of this work to investigate the consequences of this relation in detail, but we note that it is of the general form 
\begin{eqnarray} 
(\Delta_{\Psi}S_{i})^2(\Delta_{\Psi}S_{j})^2 \geq \dots \geq \left(\frac{\hbar + \beta}{2}\right)^2|(\epsilon_{ij}{}^{k})^2\braket{\hat{S}_{k}}_{\Psi}^2| \, ,
\end{eqnarray}
where the leading contribution to the terms in the middle is $(\Delta_{\Psi}\mathcal{S}_{i})^2(\Delta_{\Psi}\mathcal{S}_{j})^2 \geq (\hbar/2)^2|(\epsilon_{ij}{}^{k})^2\braket{\hat{\mathcal{S}}_{k}}_{\Psi}^2|$. 
This is equivalent to the uncertainty relation for spin measurements in canonical QM. 
The additional terms are non-canonical and depend on the ratio of the dark energy density to the Planck density, which determines the value of the geometry quantisation scale, $\beta$ \cite{Lake:2019nmn,Lake:2020chb}.

\subsection{Relation to the theory of quantum reference frames} \label{sec:4.4}

In Sec. \ref{sec:4.1} we introduced a useful unitary transformation, $\hat{U}_{\beta}$ (\ref{U_beta}). 
The action of $\hat{U}_{\beta}$ symmetrises the rigged bases of the `extended' Hilbert space, to which the smeared state $\ket{\Psi}$ belongs, such that $\ket{{\bf x},{\bf x'}} \mapsto \ket{{\bf x},{\bf x' - x}}$ (\ref{U_beta_action_pos}) and $\ket{{\bf p \,  p'}} \mapsto \ket{{\bf p},{\bf p' - p}}$ (\ref{U_beta_action_mom}). 
In the first set of bases $\ket{\Psi}$ is non-separable so that matter and geometry appear entangled, as in Eqs. (\ref{|Psi>_position_space}) and (\ref{Psi_p}). 
This is consistent with the hypothesis advanced in \cite{Kay:2018mxr}. 
However, in general, entanglement is a basis-dependent property of quantum states \cite{Chuang_Nielsen,Ish95}. 
In the symmetrised bases the smeared state is separable, $\ket{\Psi} = \ket{\psi} \otimes \ket{g}$ (\ref{Psi_unitary_equiv}), so that matter and geometry are no longer entangled. 

It follows that, if $\hat{U}_{\beta}$ represents a viable physical transformation of the system, then the entanglement of matter and geometry in the smeared space model is frame dependent.
This is consistent with recently obtained results in the theory of quantum reference frame (QRF) transformations \cite{Giacomini:2017zju}. 
In fact, there exists a formal similarity between Eq. (\ref{Psi_unitary_equiv}) and the separable pre-QRF state considered in \cite{Giacomini:2017zju}, and between Eq. (\ref{|Psi>_position_space}) and the entangled post-QRF state considered therein. 
The unitary operator $\hat{U}_{\beta}$ (\ref{U_beta}) is formally analogous to the operator that switches between QRFs in this formalism \cite{Lake:2019nmn}. 
We recall that, in \cite{Giacomini:2017zju}, a QRF is defined as a superposition of classical reference frames. 
This is intended to represent the realistic `reference frame' defined by an observer that is embodied as a quantum system.  

The considerations above suggest a link between QRFs and GURs. 
Roughly speaking, we may imagine the entangled matter-plus-geometry state of the first formalism, Eq. (\ref{|Psi>_position_space}), as the state of a QM particle `seen' by an observer embodied as a quantum spatial point, $\ket{{\bf x}} \otimes \ket{g_{{\bf x}}}$ (\ref{smearing_map}). 
This explains the primary difference between the states and transformations considered in the smeared space model and their counterparts in the QRF theory  \cite{Giacomini:2017zju,Lake:2019nmn}. 
It may be verified that Eqs. (\ref{|Psi>_position_space}), (\ref{Psi_p}) and (\ref{U_beta}) reduce to their counterparts in \cite{Giacomini:2017zju} in the limit $\hbar \rightarrow \beta$. 
This is consistent with the fact that the QRF formalism describes observers embodied as quantum systems on a fixed classical background whereas the smeared space formalism treats the background itself as a quantum mechanical object \cite{Lake:2019nmn}. 
\footnote{In \cite{Lake:2019nmn}, it was argued that the operator $\hat{U}_{\beta}$ does not represent a physically accessible transformation of the composite matter-plus-geometry system. However, even if the entanglement of matter and geometry is a frame-independent feature of the theory, the formal similarity between the smeared space formalism and the formalism of QRF transformations, under the exchange $\beta \rightarrow \hbar$, still holds. This strongly suggests that our interpretation of the smeared state (\ref{|Psi>_position_space}), given above, remains valid. It may be hoped that future work can either confirm or disprove this \cite{LakeMiller2020}.} 

However, on reflection, there is clearly something missing from both formalisms. 
Specifically, the QRF formalism \cite{Giacomini:2017zju} allows us to describe the results of physical experiments, as seen by observers that are embodied as canonical quantum systems in a fixed classical background geometry. 
By contrast, the smeared space formalism allows us to describe the results of experiments, as seen by observers embodied as quantum spatial points \cite{Lake:2018zeg,Lake:2019nmn}. 
In any viable theory of quantum gravity, we expect realistic observers to be embodied as material quantum systems living on, or `in', a quantum background geometry. 
This suggests that the two formalisms could be combined to give a deeper picture of this scenario. 

In principle, the combined theory should allow us to describe superpositions of uniformly accelerated reference frames (i.e., uniform gravitational fields), in which each Planck-sized volume of space is subject to additional quantum fluctuations, as expected from generic arguments in phenomenological quantum gravity \cite{Crowell:2005ax,Garay:1994en,Hossenfelder:2012jw,Padmanabhan:1985jq}. 
At present, the QRF formalism is able to describe the former, but not the latter, whereas the smeared space formalism describes the latter, but not the former. 
It may be hoped that future research will yield further insights into the nature of both QRFs and GURs, and the possible connections between them \cite{LakeMiller2020}. 

\section{Avoiding the conventional wisdom} \label{sec:5}

In this section, we consider possible objections to the smearing procedure presented in \cite{Lake:2018zeg,Lake:2019nmn}. 
Two problems come immediately to mind. 
Both are potentially valid and must be addressed head on, if the model is to survive as a viable competitor to the standard modified commutator theories. 
Due to limitations of space, our discussion is brief, and the problems cannot be considered as fully resolved. 
Nonethless, there are encouraging signs that they may be overcome within the existing formalism, and it may be hoped that future research will be able to settle these issues one way or another.  

The first problem is that there exist well known no go theorems for multiple quantisation constants (see, for example, \cite{Sahoo2004} and references therein). 
In Sec. \ref{sec:5.1}, we consider whether the existing theorems apply to the smeared states defined in Secs. \ref{sec:3.1}-\ref{sec:3.2} and \ref{sec:4.1}, and argue that they do not. 
The key point is that the smeared space formalism treats matter and geometry asymmetrically. 
This is expressed, mathematically, via the modified de Broglie relation (\ref{mod_dB*}), in which ${\bf k}$ and ${\bf k'}$ represent independent degrees of freedom, but ${\bf p'}$ remains the only observable momentum. 
Therefore, quantising the composite matter-plus-geometry system is not equivalent to quantising the state of two, distinguishable, material particles. 
Because of this, the standard theorems cannot be applied, directly, to states in the smeared space model. 

The second problem is that, as every quantum gravity researcher knows, canonical quantisation of the `metric' yields a spin-2 representation of the inhomogeneous Lorentz group \cite{Pauli-Fierz,Wigner:1939cj}. 
At first sight, this is immediately at odds with our claim, in Sec. \ref{sec:4.3}, that the fundamental quanta of space are fermions. 
In Sec. \ref{sec:5.1}, we present a critical view of the canonical quantisation procedure \cite{Pauli-Fierz} and argue that the standard interpretation applies only to the first order perturbation of the metric. 
Crucially, the leading order term, which represents the flat background space whose symmetries are defined by the Poincar{\'e} group \cite{SRFrench}, remains purely classical in the usual quantisation of linearised gravity \cite{Pauli-Fierz}. 
We emphasise that the new degrees of freedom introduced in the smeared space formalism correspond to the non-relativistic limit of this flat piece, yielding superpositions of Euclidean geometries \cite{Lake:2018zeg,Lake:2019nmn}. 
Therefore, our proposal does not contradict existing results about the spin-2 nature of gravitons, which are represented by the quantised perturbation.  

\subsection{Evading no go theorems for multiple quantisation constants} \label{sec:5.1}

Well known no go theorems forbid the canonical quantisation of different material particle species using different quantisation constants \cite{Barcelo:2012ja,Gil:2016beb,Sahoo2004}. 
Given two distinguishable particles, A and B, whose degrees of freedom are labelled by primed and unprimed variables, respectively, it may be shown that imposing the canonical-type de Broglie relations
\begin{eqnarray} \label{h_h'}
{\bf p} = \hbar{\bf k} \, , \quad {\bf p'} = \hbar'{\bf k'} \, ,
\end{eqnarray} 
with $\hbar \neq \hbar'$, leads to fundamental inconsistencies \cite{Sahoo2004}. 

In the limit $\hbar' \rightarrow \hbar$, the relations (\ref{h_h'}) reduce to the standard ones, ${\bf p} = \hbar{\bf k}$ and ${\bf p'} = \hbar{\bf k'}$. 
We simply note that, in the limit  $\beta \rightarrow \hbar$, the quantisation conditions of the smeared space formalism, ${\bf p} = \hbar{\bf k}$ and ${\bf p'-p} = \beta({\bf k'-k})$ (\ref{mod_dB}), which combine to give ${\bf p'} = \hbar{\bf k} + \beta({\bf k'-k})$ (\ref{mod_dB*}), also reduce to the standard relations. 
However, crucially, setting $\beta = \hbar' \neq \hbar$ does not recover Eqs. (\ref{h_h'}). 
For this reason, the usual no go theorems presented in \cite{Barcelo:2012ja,Gil:2016beb,Sahoo2004}, and in related literature, are not directly applicable to states in the smeared space model. 
The existence of such states, given by Eqs. (\ref{|Psi>_position_space}), (\ref{Psi_p}) and (\ref{Psi_unitary_equiv}), is not expressly forbidden by these theorems. 

Nonetheless, it is of course concievable that even slight modifications of the existing theorems could be used to demonstrate inconsistencies in the smearing procedure, represented by Eqs. (\ref{smearing_map}) and (\ref{mod_dB}). 
This possibility must be thoroughly investigated, before the proposal of a second quantisation scale for space can be taken seriously \cite{Lake:2020chb}. 
This work is of the utmost importance for the future development of the smeared space model. 
If inconsistencies can be explicitly demonstrated, it must be abandoned as a physical theory. 

\subsection{Isn't spacetime spin-$2$? Gravitons versus quanta of space} \label{sec:5.2}

The standard argument for the spin-$2$ nature of spacetime is that, since the classical metric is a symmetric two-index tensor field, its canonical quantisation yields a spin-$2$ representation of the inhomogeneous Lorentz group in the tangent space associated with each spacetime point. 
This argument is very robust as it forms part of a general scheme that may be used to construct the dynamical equations for particles of any spin \cite{LieAlgebrasGutowski,Jones98} including photons, spin-$1/2$ fermions, and even the composite spin-$3/2$ particles predicted by the Standard Model \cite{Donoghue:1992dd,KramersMethod}. 
However, on closer inspection, a linguistic sleight of hand has been applied with the use of the word `metric'.

In the standard approach the full metric of general relatively, $g_{\mu\nu}$, is expanded to first order:
\begin{eqnarray} \label{metric_pert}
g_{\mu\nu} \simeq \eta_{\mu\nu} + h_{\mu\nu} \, . 
\end{eqnarray}
Thereafter, $\eta_{\mu\nu}$ and $h_{\mu\nu}$ are regarded as dynamical and non-dynamical pieces, respectively. 
This is expressed by the fact that $\eta^{\mu\nu} = (\eta_{\mu\nu})^{-1}$ and $\eta_{\mu\nu}$ are used to raise and lower indices on $h_{\mu\nu}$ and $h^{\mu}{}_{\nu}$, etc., as well as on fields of any other type, corresponding to various forms of matter. 

The metric perturbation is therefore defined as a field on flat spacetime \cite{Pauli-Fierz}. 
This enables canonical quantisation techniques, which are valid for fields in Minkowski space, to be applied to $h_{\mu\nu}$. 
These are not applicable to $g_{\mu\nu}$ \cite{DeWittMorette:2011zz}. 
The resulting equations of motion, known as the Pauli-Fierz equations, describe the dynamics of spin-$2$ particles, aka gravitons \cite{Pauli-Fierz}:
\begin{eqnarray} \label{Pauli-Fierz}
\Box h_{\mu\nu} = 0 \, , \quad \partial_{\nu} h_{\mu\nu} = 0 \, , \quad h^{\mu}{}_{\mu} = 0 \, .
\end{eqnarray}
Hence, only the quantized perturbation of the metric yields a spin-2 representation of the inhomogeneous Lorentz group, which is a subgroup of the Poincar{\' e} group that defines the symmetries of flat Minkowski space on which it `lives' \cite{SRFrench,Jones98}. 
The Minkowski piece of the expansion (\ref{metric_pert}) remains non-dynamical and completely classical.

In the non-relativistic limit a similar split yields a classical Euclidean background in which the perturbation $h_{00}$ is associated with the Newtonian gravitational potential and the $h_{ij}$ components are neglected \cite{Hobson:2006se}. 
We stress that the fermionic quanta of space predicted by our model are associated with the flat Euclidean piece of the metric, which remains classical and non-dynamical in the standard approach. 
By contrast, we endow the Euclidean background with a quantum genuine state vector in the Hilbert space of the theory \cite{Lake:2018zeg,Lake:2019nmn}. 
This forms part of the composite matter-plus-geometry state that obeys a modified Schr{\" o}dinger equation (see \cite{Lake:2018zeg} for details).

Though many aspects of the smeared space theory remain unclear, and have yet to be explored, the qualitative picture that emerges is of fermionic spacetime quanta exchanging virtual gravitons, in addition to the usual graviton-matter and graviton-graviton interactions \cite{Lake:2018zeg,Lake:2019nmn}. 
This is perhaps not so crazy. 
We recall that in the Standard Model of particle physics the fundamental constituents of matter are fermions, which exchange virtual bosons as force-mediators \cite{Donoghue:1992dd,Peskin:1995ev}. 
Here, we treat gravitons as the fundamental quanta of curvature, which give rise to the quantised gravitational force, but model the underlying spacetime fabric as a web of fermions. 
This has the major advantage of allowing us to answer the following question, which cannot be addressed within the standard paradigm of linearised quantum gravity: of what is the quantum space composed when it is flat?

\section{Discussion} \label{sec:6}

We have presented a new approach to GURs, based on the `smearing' of classical points into coherent superpositions of point-like quantum states. 
The mathematical formalism of the model was presented in two forms. 
The first incorporates generalised position and momentum measurements, which naturally generate the EGUP (\ref{smeared-spaceEGUP-2})-(\ref{smeared-spaceEGUP-3}), while the second allows us to define generalised operators for angular momentum and spin, giving rise to GURs for these observables. 
The key advantage of the new approach is that it generates GURs without violating the symmetry algebras of canonical QM. 
This allows it to evade the well known problems associated with modified commutation relations, which are assumed to form the mathematical basis of GURs in the bulk of the existing literature \cite{Hossenfelder:2012jw,Kempf:1994su,Nicolini:2012eu,Tawfik:2015rva,Tawfik:2014zca}. 

However, this phenomenological success comes at a price and the model implies two radical conclusions for which there is, at present, no empirical support. 
The first is that space, and hence gravity, must be quantised at a different scale to matter, $\beta \simeq \hbar \times 10^{-61}$ (\ref{beta_mag}) \cite{Lake:2018zeg,Lake:2020chb}. 
The second is that the fundamental quanta of the space-time fabric are fermions, with spin $\pm \beta/2$, (\ref{unentangled_eigenvectors})-(\ref{entangled_eigenvectors}) \cite{Lake:2019nmn}. 
To conclude, we give a critical appraisal of the current status of the model, highlighting what is missing from the present formalism, and what open questions need still to be addressed. 

\subsection{Current status of the model} \label{sec:6.1}

A valid criticism of the current formalism is that the smearing scales for position and momentum, $\Delta_{g} x'^i \simeq l_{\rm Pl} \simeq 10^{-33} \, {\rm cm}$ and $\Delta_{g} p'_i \simeq m_{\rm dS}c \simeq 10^{-56} \, {\rm g \, . \, cm \, s^{-1}}$, are put into the theory by hand. 
That is, although the wave-point uncertainty relation (\ref{Beta_UP}) bounds the uncertainties associated with a quantum spatial `point' such that $\Delta_{g} x'^i \Delta_{g} p'_i \gtrsim (3/2)\beta$, where $\beta \simeq l_{\rm Pl}m_{\rm dS}c \simeq \hbar \times 10^{-61}$ (\ref{beta_mag}), this does not bound the values of $\Delta_{g} x'^i$ or $\Delta_{g} p'_i$ individually. 

Because (\ref{Beta_UP}) is formally analogous to the HUP (\ref{HUP-2}), under the exchanges $g \leftrightarrow \psi$ and $\beta \leftrightarrow \hbar$, it is consistent with the limits $\Delta_{g} x'^i \rightarrow 0$, $\Delta_{g} p'_i \rightarrow \infty$ and $\Delta_{g} x'^i \rightarrow \infty$, $\Delta_{g} p'_i \rightarrow 0$, respectively, so long as $\beta$ remains finite. 
However, in the former, the minimum momentum uncertainty of a particle in the smeared background space also tends to infinity, whereas, in the latter, the minimum position uncertainty diverges \cite{Lake:2018zeg}. 
Only by assuming the fixed values, given above, can the formalism give rise to the EGUP expected from model-independent gedanken experiments (\ref{smeared-spaceEGUP-2})-(\ref{smeared-spaceEGUP-3}) \cite{Lake:2018zeg,Lake:2019nmn}.  

The non-dynamical nature of the smearing functions, $g({\bf x'-x})$ and $\tilde{g}_{\beta}({\bf p'-p})$, is therefore introduced as a hypothesis of the model. 
At its present level of development, this seems unavoidable. 
Nonetheless, it would certainly be preferable to derive these conditions from a more fundamental dynamical theory. 
Though speculative, we conjecture that an appropriate generalisation of the Pauli exclusion principle (PEP), incorporating the spin-spin interaction between matter and geometry, may enable us to achieve this. 
The rough idea is that the fermionic degeneracy pressure between delocalised spatial `points' is precisely what keeps them delocalised, ensuring that a unique function $\braket{{\bf x'}|g_{{\bf x}}} = g({\bf x'-x})$ exists for each value of ${\bf x}$. 

Whether or not this speculation turns out to be true, it is clear that the canonical PEP must be generalised in the context of the smeared space model \cite{Lake:2019nmn}. 
In fact, virtually all the predictions of canonical QM could (and should) be generalised to incorporate the effects of the smeared background. 
These include all the more recent developments of the theory, conventionally grouped under the umbrella of quantum information theory (QIT) \cite{Chuang_Nielsen}. 
There is lots of work still to do, even in the non-relativistic limit. 
Furthermore, we must now think seriously about why, and how, the theory should be extended into the relativistic regime, and about how to introduce gravity into the model in a more fundamental way. 
In the final section, \ref{sec:6.2}, we briefly consider these issues, and outline various prospects for future work.

\subsection{Future work} \label{sec:6.2}

Ultimately, we would like to construct a smeared version of the Standard Model of particle physics, including all known interactions of the electromagentic, weak, and strong nuclear forces \cite{Donoghue:1992dd}. 
Outstanding problems for this research program include the questions of how to `smear' time, how to smear general gauge symmetries, and how to formulate the corresponding path integral picture \cite{Lake:2018zeg,Lake:2019nmn}. 
However, even if we were able to overcome each of these problems, our model still would not contain a fundamental description of gravity. 

This is ironic, for a model that was originally motivated by gedanken experiments in quantum gravity phenomenology \cite{Adler:1999bu,Bambi:2007ty,Bolen:2004sq,Maggiore:1993rv,Park:2007az,Scardigli:1999jh}. 
Nonetheless, it is consistent with the simplest picture of Newtonian gravity, in which the gravitational potential is viewed as a scalar field in flat Euclidean space \cite{Hobson:2006se,Lake:2018zeg,Lake:2019nmn}. 
To incorporate gravity on a more fundamental level, two options immediately suggest themselves. 
The first is to complete the research program outlined above by extending the smearing procedure to flat Minkowski space and the Standard Model of particles physics. 
Armed with a superposition of flat spacetimes, we may then consider how to construct superpositions of curved spacetime geometries \cite{Lake:2018zeg}. 
The second is to incorporate Newtonian gravity into the model in a more fundamental way. 
In principle, this may be done by smearing the Newton-Cartan formalism, in which non-relativistic gravity emerges from the curvature of bona fide Riemmanian geometries \cite{Banerjee:2018gqz,Hansen:2018ofj}.

Thankfully, these approaches are not mutually exclusive, and we may pursue both simultaneously. 
In fact, this is a good strategy. 
If gravity is to be viewed as spacetime curvature then the current smeared space theory corresponds, at best, to the $G \rightarrow 0$, $c \rightarrow \infty$ limit of the complete quantum gravity theory. 
By contrast, the generalisations proposed above correspond to different limits, namely, $G \rightarrow 0$ with finite positive $c$, and $c \rightarrow \infty$ with finite positive $G$, respectively. 
Constructing each of these limits, explicitly, gives us two possible routes by which to attack the fundamental problem of relativistic quantum gravity \cite{Lake:2018zeg}. 

Finally, the model should be probed, at every stage of its development, for mathematical inconsistencies. 
(For example, those associated with multiple quantisation constants, as discussed in Sec. \ref{sec:5.1}.) 
If possible, experimental schemes should also be derived to test it empirically. 
Though it seems unlikely that table-top experiments could ever be sensitive enough to directly probe the smeared geometry, considering the extraordinarily small value of $\beta$, this cannot be discounted a priori. 
In addition, we may explore the phenomenological implications of the model for present day cosmology and the history of the universe \cite{Lake:2018zeg,Lake:2019nmn}. 
This may prove to be an especially fruitful research direction, given the intimate connection between $\beta$ and the observed vacuum energy, $\rho_{\Lambda}$ (\ref{beta}) \cite{Lake:2020chb}. 

\section*{Acknowledgements}
I am extremely grateful to Marek Miller for helpful comments, suggestions, and discussions. 
My thanks also to Piero Nicolini, for sharing his understanding of nonlocal geometry models, and to Michael Hall, for bringing several references to my attention. 
This work was supported by the Natural Science Foundation of Guangdong Province, grant no. 008120251030.




\end{document}